\documentclass[useAMS,usenatbib]{mn2e}

\usepackage{amssymb}
\usepackage{mathptmx}       
\usepackage{helvet}         
\usepackage{courier}        
\usepackage{type1cm}        
 \usepackage{makeidx}         
\usepackage{graphicx}        
\usepackage{multicol}        
\usepackage[bottom]{footmisc}

\usepackage{natbib}
\usepackage{aas_macros}  

\usepackage{color}
\usepackage[usenames,dvipsnames]{xcolor}
\usepackage{textcomp}
\usepackage{times,deluxetable}

\title[Virgo cluster and field dwarf ellipticals in 3D: III.]{Virgo cluster and field dwarf ellipticals in 3D: \\III. Spatially and temporally resolved stellar populations}

\author[Agnieszka Ry\'{s} et al.]{Agnieszka Ry\'{s}$^{1,2}$\thanks{E-mail: arys@eso.org}, Mina Koleva$^{3}$, Jes\'{u}s Falc\'{o}n-Barroso$^{2,4}$, Alexandre Vazdekis$^{2,4}$, 
\newauthor Thorsten Lisker$^{5}$, Reynier Peletier$^{6}$, Glenn van de Ven$^{7}$\\ 
$^{1}$European Southern Observatory, Karl-Schwarzschild-Strasse 2, 85748 Garching bei Muenchen, Germany\\
$^{2}$Instituto de Astrofísica de Canarias, V\'ia L\'actea s/n, E-38205 La Laguna, Tenerife, Spain\\
$^{3}$Sterrenkundig Observatorium, Ghent University, Krijgslaan 281, S9, B-9000 Gent, Belgium\\
$^{4}$Departamento de Astrof\'isica, Universidad de La Laguna, E-38205 La Laguna, Tenerife, Spain\\
$^{5}$Astronomisches Rechen-Institut, Zentrum f\"{u}r Astronomie der Universit\"{a}t Heidelberg, M\"{o}nchhofstrasse 12-14, D-69120 Heidelberg, Germany\\
$^{6}$Kapteyn Astronomical Institute, University of Groningen, Postbus 800, 9700 AV Groningen, the Netherlands\\
$^{7}$Max Planck Institute for Astronomy, K\"{o}nigstuhl 17, 69117 Heidelberg, Germany
}

\begin{document}
\label{firstpage}
\maketitle

\begin{abstract}
We present the stellar population analysis of a sample of 12 dwarf elliptical galaxies, observed with the SAURON integral field unit, using the full-spectrum fitting method. We show that star formation histories (SFHs) resolved into two populations can be recovered even within a  limited wavelength range, provided that high signal-to-noise data is used. We confirm that dEs have had complex SFHs, with star formation extending to (more) recent epochs: for the majority of our galaxies star formation activity was either still strong a few ($\lesssim$\,5) Gyr  ago or they experienced a secondary burst of star formation roughly at that time. This latter possibility is in agreement with the proposed dE formation scenario where tidal harassment drives the gas remaining in their progenitors inwards and induces a star formation episode. For one of our field galaxies, ID\,0918, we find a correlation between its stellar population and kinematic properties, pointing to a possible merger origin of its kinematically-decoupled core. One of our cluster objects, VCC\,1431, appears to be composed exclusively of an old population ($\gtrsim$\,10-12\,Gyr). Combining this with our earlier dynamical results, we conclude that the galaxy was either ram-pressure stripped early on in its evolution in a group environment and subsequently tidally heated, or that it evolved \textit{in situ} in the cluster's central parts, compact enough to avoid tidal disruption. These are only two of the examples illustrating the SFH richness of these objects confirmed with our data.

\end{abstract}

\begin{keywords}
galaxies: dwarf -- galaxies: evolution -- galaxies: formation -- galaxies: stellar populations -- galaxies: structure
\end{keywords}

\section{Introduction}

Dwarf elliptical galaxies (dEs) have, as a class, received a significant amount of attention in the literature in recent years. This is because with the advent of higher resolution/more sensitive instruments it became feasible to study these low-surface-brightness systems in more detail. A great deal of complexity of their structure has been revealed, prompting some works to introduce a complex classification system and sometimes view the different subclasses as intrinsically distinct populations. We know that their photometry shows underlying structures such as spiral arms (e.g. \citealt{jerjen:2000}), bars, and disks (e.g. \citealt{lisker:2006a}a). A similar variety was found in their kinematic and stellar population properties. We know of Virgo and field dEs harboring kinematically-decoupled cores (e.g. \citealt{derijcke:2004}, \citealt{geha:2005}, \citealt{rys:2013}, \citealt{toloba:2014}, \citealt{guerou:2015}). dE stellar populations show indications of both young and old ages and varied gradients (e.g. \citealt{koleva:2009}, \citealt{koleva:2011}, \citealt{rys:2012jenam}).

Recent evidence has caused the majority of literature to favor the environmental transformation scenario for dEs , where their progenitors are assumed to be late-type galaxies that were transformed once they entered the denser (cluster) environment (e.g.  \citealt{geha:2010}, \citealt{janz:2012}, \citealt{kormendy:2012}, \citealt{toloba:2012}, \citealt{lisker:2013}, \citealt{rys:2013}). The notion was, in fact, first proposed as early as 1944 by \citeauthor{baade:1944} based on his study of Local Group dEs NGC\,185 and NGC\,147. The claim reappeared later in the works of \cite{kormendy:1985}, \cite{binggeli:1988}, \cite{binggeli:1991} and \cite{bender:1992}. A few mechanisms have been proposed that may be responsible for such environmentally-induced transformation. Ram-pressure stripping (\citealt{gunn:1972}, \citealt{lin:1983}) can remove a galaxy's remaining gas reservoir on relatively short time scales so that the star formation stops quickly once the galaxy enters denser environments. It has been directly shown for massive spiral galaxies in Virgo in Chung et al. (2007). In the galaxy harassment scenario (\citealt{moore:1998}) the tidal interactions between the galaxy and the intergalactic medium (or in the extreme cases, galaxy-galaxy interactions) can heat up the object and slow its rotation down (albeit not remove it completely). They are also able to remove both stellar and, more strongly, dark mass. 

Finding the progenitors of dEs is not a straightforward task. First of all because what we look at today are \textit{present-day analogs of the presumed late-type progenitors} of dEs. Thus, none of the late-type galaxy types  considered to be progenitor candidates for dEs (e.g. dIrrs, dwarf spirals) can really play that role. They have evolved in parallel, albeit evidently under different circumstances. This was first pointed out by \cite{binggeli:1994} and \cite{skillman:1995} and repeated more recently in, e.g. \cite{lisker:2013} who also added that the conditions a few Gyr ago were necessarily different, in terms of density distribution and so the interactions between galaxies and between galaxies and the surrounding medium did not resemble those we see today.

In observational studies only general statements are typically made about certain scenarios being compatible (or not) with a given observational result. Few attempts have been made at a quantitative discrimination between the processes because this requires making a priori assumptions about the progenitors. We have faced a similar challenge in the interpretation of our previous work. In \cite{rys:2013} (Paper I of our series), the variety of properties we discuss there could be due to either stochastic environmental processes or because the progenitor family consists of galaxies of various types. Later in \cite{rys:2014}, hereafter Paper II, we compared the dynamical properties of our dEs to those of dwarf and giant late-type objects and we concluded that a transformation mechanism is required which not only is able to lower the angular momentum but also needs to account for the increased stellar concentration of dEs with respect to their presumed progenitors. This could be achieved either in the case of harassment being the important mechanism, or it could be explained by assuming that the progenitors were already more compact at higher redshift. While these findings have naturally narrowed down the range of possibilities for the formation paths, a number of open questions still remained. 

Here (Paper III of the series) we investigate the star-formation histories of our sample using the full-spectrum fitting technique, with the view to further narrowing down the above results. The paper is structured as follows. A summary of the sample selection, observations, and data reduction is presented in Section 2. In Section 3 we describe the methods used in the analysis. Section 4 presents our results which are then discussed in Section 5. The findings are summarized in Section 6..

\begin{table}
\caption{Observed objects: name, distance (for the Virgo objects from Mei et al. 2007 where available, otherwise 17 Mpc assumed), morphological type (from Lisker et al. 2007), ellipticity, r-band effective radius, and r-band apparent magnitude. Adapted from \protect\cite{rys:2013}.} 
\centering
\begin{tabular}{|r|r|r|r|r|r|}
\hline
object&distance& type   &$\epsilon$&R$_e$ & m$_r$\\ 
      & $(Mpc)$&        &          &$('')$&$(mag)$\\ 
\hline
VCC 0308 &17.00&dE(di;bc)&0.07&18.7&13.32\\
VCC 0523 &16.74&dE(di)   &0.29&27.9&12.60\\
VCC 0929 &17.00&dE(N)    &0.11&22.1&12.65\\
VCC 1036 &17.00&dE(di)   &0.56&17.2&13.13\\
VCC 1087 &16.67&dE(N)    &0.31&28.6&12.85\\
VCC 1261 &18.11&dE(N)    &0.42&19.7&12.87\\
VCC 1431 &17.00&dE(N)    &0.03& 9.6&13.60\\
VCC 1861 &16.14&dE(N)    &0.01&20.1&13.41\\
VCC 2048 &17.00&dE(di)   &0.48&16.5&13.08\\
NGC 3073 &17.8 &dE/dS0 &0.15&16.1&12.98\\
ID 0650  &25.9 &dE/S0     &0.10&20.1&13.73\\
ID 0918  &16.3 &dE/E       &0.27& 6.4&13.79\\
\hline
\end{tabular}
\label{observations} 
\end{table}

\begin{figure*}
\begin{centering}
\hspace{2cm}\includegraphics[width=0.29\textwidth,angle=90]{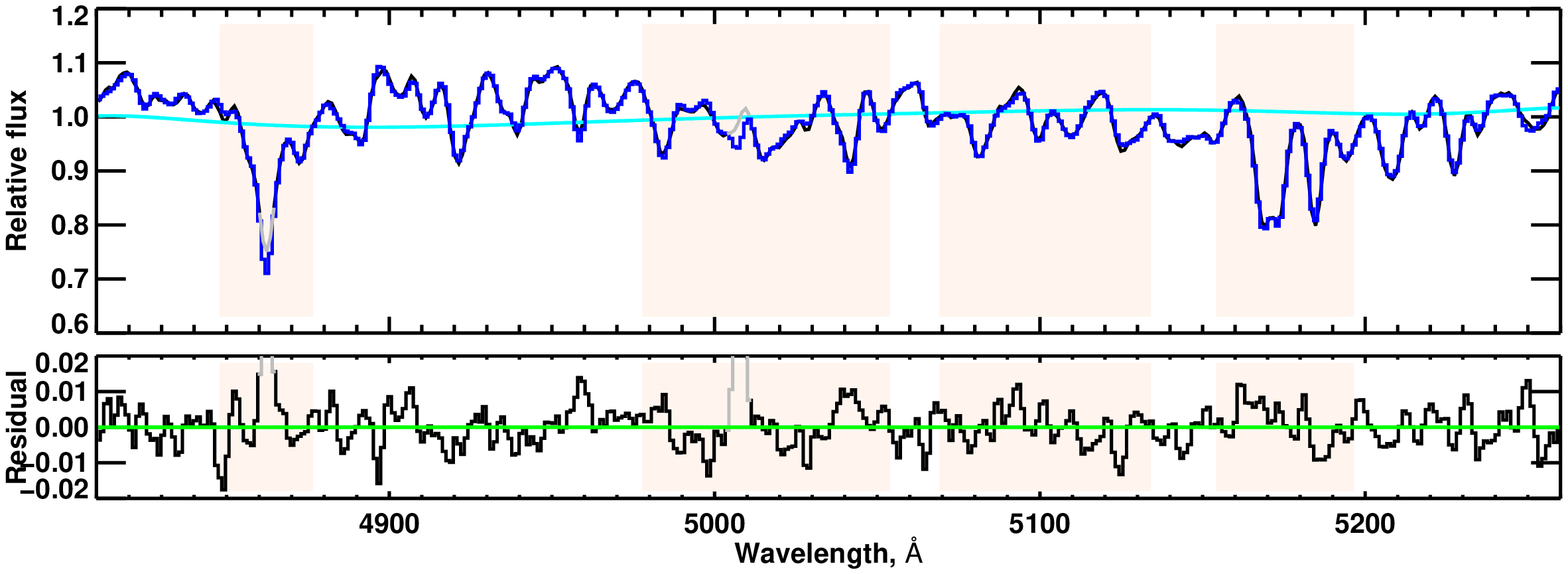}\\
\hspace{2cm}\includegraphics[width=0.29\textwidth,angle=90]{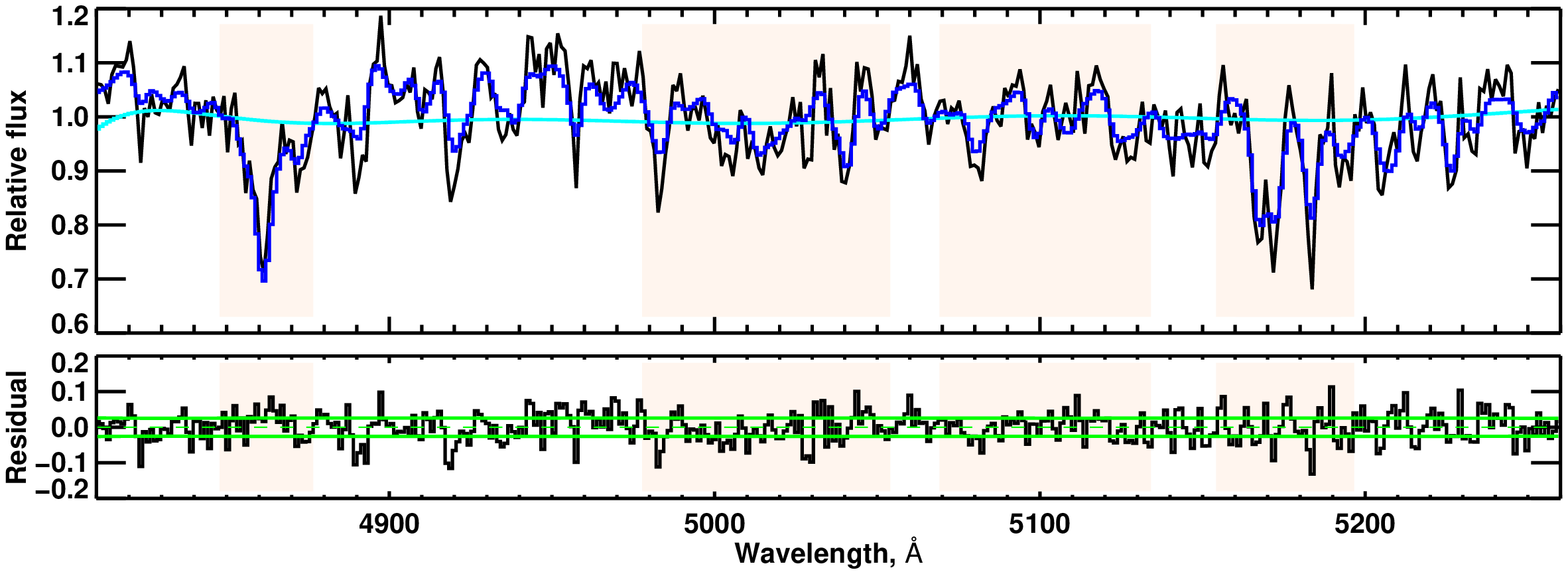}

\end{centering}
\caption{Example ULySS fits to the VCC\,0523 galaxy central spectrum (collapsed within an aperture with a 3-arcsec diameter) for our SAURON data (upper panel) and SDSS data (lower panel), shown here for the SAURON wavelength range. The ordinate values are plotted on normalized scales. The S/N of the SAURON spectra exceeds 100 per pixel. The data are shown with black lines, models -- blue lines, and the masked pixels are in grey. The light blue line is the multiplicative polynomial used in the fit. The regions of Lick indices are shaded in light pink. In the small panels below the spectrum we plot the residuals between the data and the model, shown as a percentage of the original value (note that in the case of SDSS the y-range of the bottom panels is 10 times larger than that for the SAURON spectra).}

\label{example-spectrum}  
\end{figure*}

\section[Data]{Data}

Details on data selection, observations and data reduction are presented in \cite{rys:2013}. In short, we observed 12 dEs: nine in the Virgo Cluster and three in the field. Our objects span a wide range of ellipticities and distances from the cluster's center. The observations were carried out in Jan 2010 and Apr 2011 (8 nights in total) using the WHT/SAURON instrument at the Roque de los Muchachos Observatory in La Palma, with each galaxy typically exposed for 5h. For the extraction and calibration of the data we followed the procedures described in \cite{bacon:2001} using the specifically designed XSAURON software developed at the Centre de Recherche Astrophysique de Lyon (CRAL). We filtered out all individual spaxels with signal-to-noise (S/N) ratio $<$\,7 which roughly corresponds to surface brightness of $\mu_V\approx23.5$\,mag at the edge of the fields (slightly depending on the object)\footnote{By filtering out low-S/N spaxels we made sure that our final binned spectra were not contaminated by low-quality measurements. Simply adding up the signal of individual spaxels to form larger bins will lead to lower overall S/N if some  of the input spaxels have low S/N, and, in our case, showed as a spread of $\sigma$ values that exceeded the nominal measurement errors.}. In the original analysis presented in Papers I and II the data were spatially binned to achieve the minimum S/N ratio of 30. Here we binned the data in annuli of increasing width so as to achieve yet higher S/N ratios. Also, the properties derived in such a way are not biased towards any particular galaxy axis which presents an advantage over the traditionally employed long-slit data. It is also important to stress the full or nearly full 2D spatial coverage of our data: for the shown radial extent the coverage is $\gtrsim$95\% for all galaxies and annuli, with only one exception of the outermost annulus of VCC\,0929 ($\sim$70\%).

Figure~\ref{example-spectrum} shows an example of a central spectrum of one of the galaxies, collapsed within an aperture with a 3-arcsec diameter, used in the SDSS/SAURON comparison presented in Section 4.

\begin{figure*}
\begin{centering}
\includegraphics[width=0.7\textwidth]{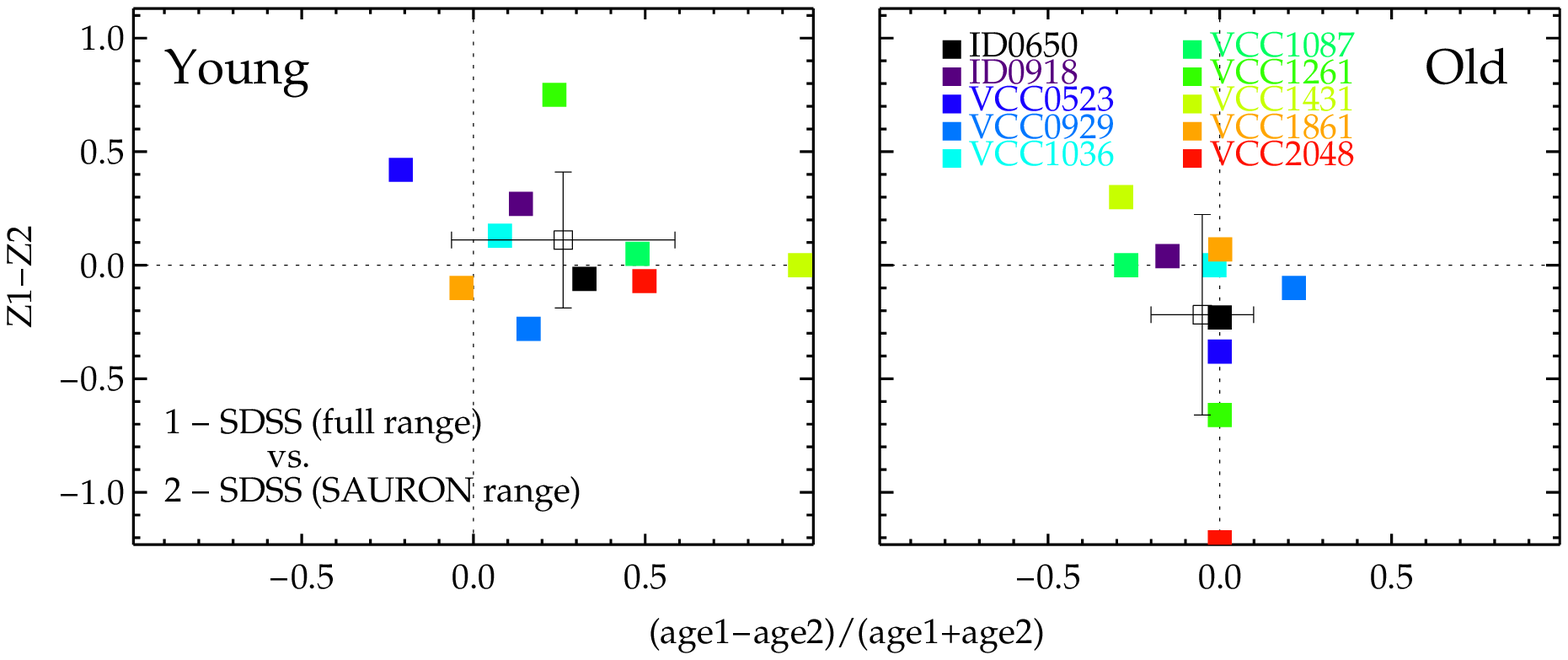}
\includegraphics[width=0.7\textwidth]{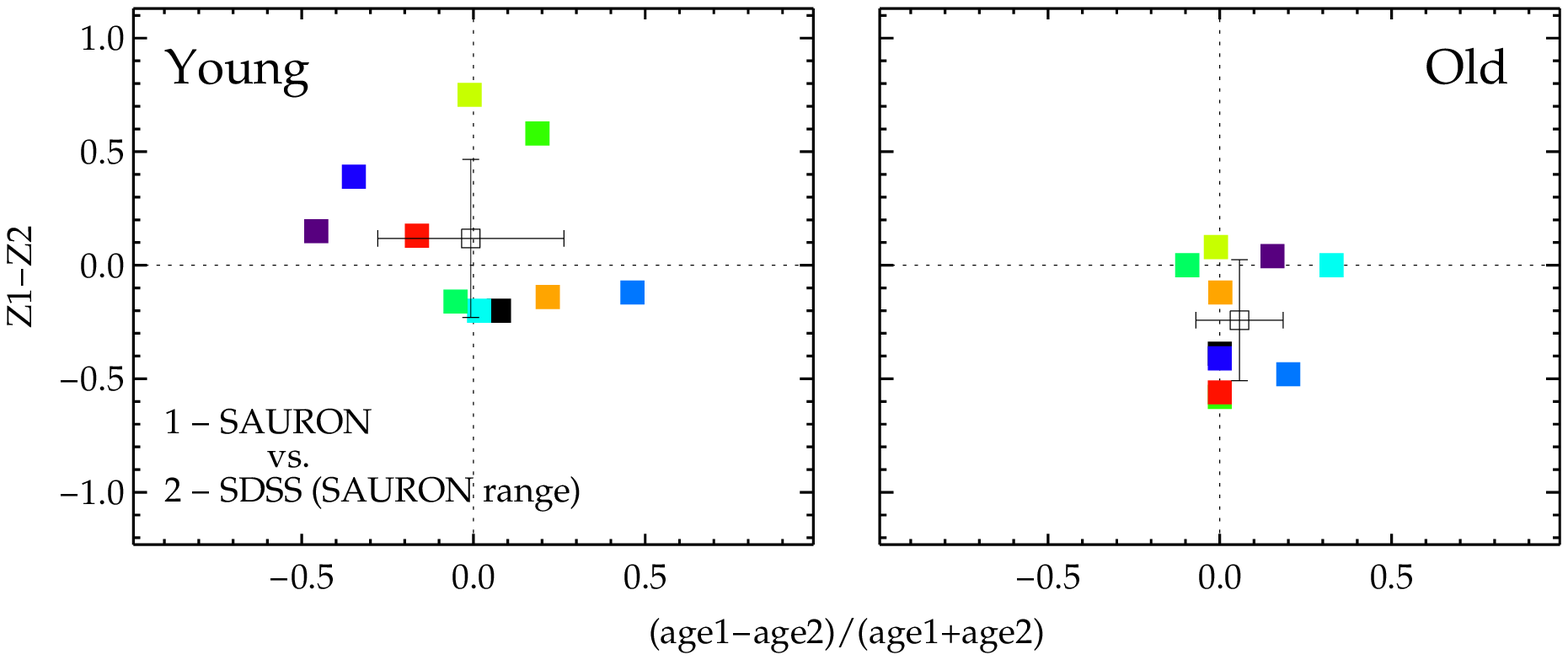}
\includegraphics[width=0.7\textwidth]{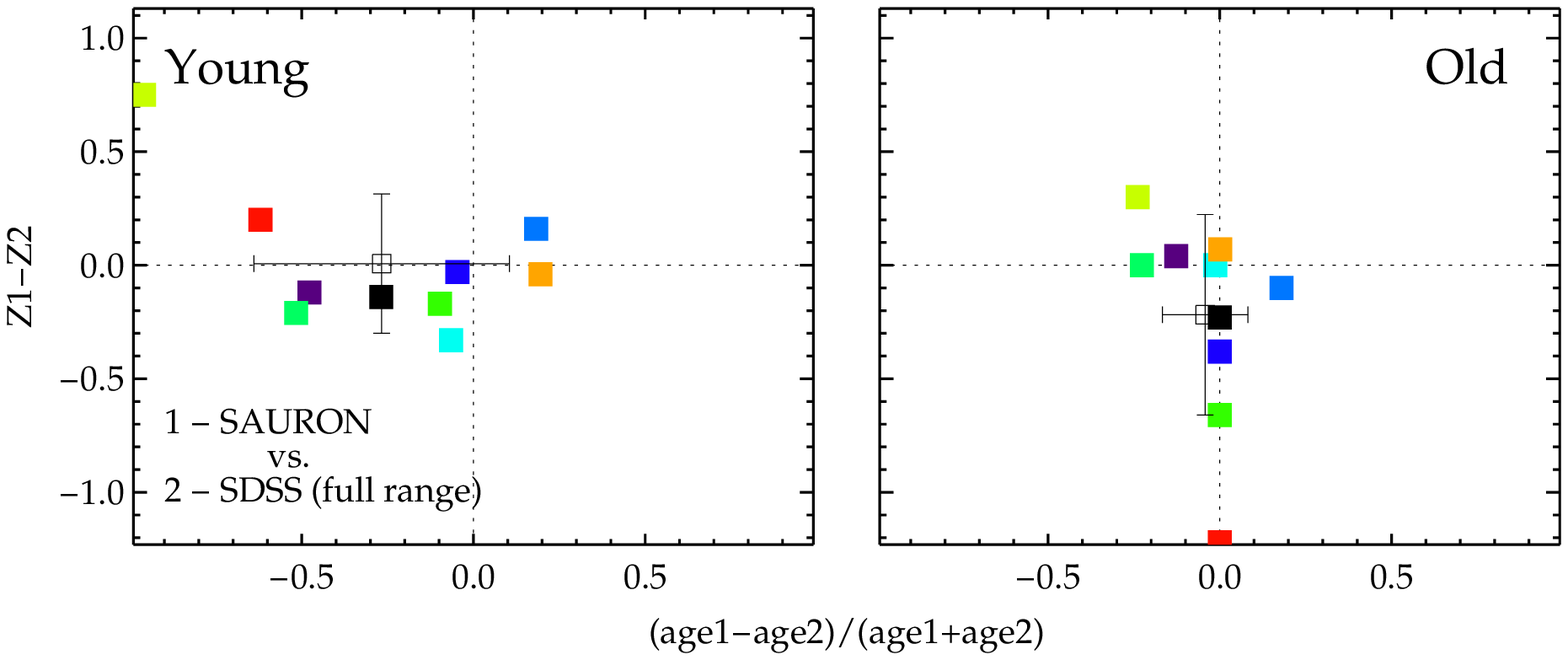}
\end{centering}
\caption{Comparison of the properties of young and old populations (ages, metallicities, and population weights from the following  data sets: 1) SDSS -- full wavelength range, 2) SDSS -- wavelength range cut to the range of SAURON, 3) central 3 arcsec of our SAURON data. The x axes show relative differences in age, and the y axes show direct differences in metallicity (Z). The dotted lines indicate where the agreement between the data sets would be perfect. The open black symbols show average values of (Z1-Z2) and (age1-age2)/(age1+age2), with the error bars indicating standard deviations. All individual errors are shown in Table~\ref{sdss-sauron-comparison-table}, which also includes population weights.}
\label{sfh-comparison}  
\end{figure*}

\begin{table*}
\caption{Tabulated results of the comparison between the SFHs extracted from the SDSS and central SAURON data. For each galaxy the first row shows the results from the fitting of full SDSS wavelength range, the second -- SDSS restricted to the wavelength range of SAURON, and the third -- central 3 arcsec of our SAURON data. Note that for some galaxies the age of the old population was fixed to 12\,Gyr, hence the 0.00 errors.} 
\centering
\begin{tabular}{|r|r|r|r|r|r|r|}
\hline
object& age (young)& Z (young)   & weight(young) & age (old) & Z (old) & weight (old)\\ 
       & (Gyr)     & (Z$_\odot$)  &               & (Gyr)     &(Z$_\odot$)&             \\ 
\hline

ID0650&  3.1$\pm$  0.4&  0.3$\pm$  0.2& 0.57& 12.0$\pm$  0.0& -1.0$\pm$  0.3& 0.43 \\
       &  1.3$\pm$  0.7&  0.3$\pm$  0.4& 0.15& 12.0$\pm$  0.0& -0.6$\pm$  0.3& 0.85 \\
       &  1.7$\pm$  0.1&  0.3$\pm$  0.1& 0.36& 12.0$\pm$  0.0& -1.3$\pm$  0.1& 0.64 \\
\hline
ID0918&  5.0$\pm$  0.0&  0.0$\pm$  0.1& 0.86& 12.3$\pm$  4.1& -1.3$\pm$  0.2& 0.14 \\
       &  1.2$\pm$  0.4&  0.4$\pm$  0.3& 0.05& 16.6$\pm$  2.9& -0.4$\pm$  0.0& 0.95 \\
       &  5.0$\pm$  0.0& -0.1$\pm$  0.0& 0.94&  5.0$\pm$  0.0& -2.3$\pm$  0.0& 0.06 \\
\hline
VCC0523&  1.0$\pm$  0.0&  0.5$\pm$  0.0& 0.21& 12.0$\pm$  0.0& -0.8$\pm$  0.1& 0.79 \\
       &  1.9$\pm$  0.3&  0.1$\pm$  0.0& 1.00& 12.0$\pm$  0.0& -0.4$\pm$  0.5& 0.00 \\
       &  0.8$\pm$  0.0&  0.4$\pm$  0.0& 0.08& 12.0$\pm$  0.0& -0.8$\pm$  0.0& 0.92 \\
\hline
VCC0929&  5.0$\pm$  0.0& -0.1$\pm$  0.1& 0.95&  9.5$\pm$  8.6& -2.2$\pm$  0.7& 0.05 \\
       &  2.1$\pm$  0.3&  0.7$\pm$  0.0& 0.16& 10.4$\pm$  2.5& -0.6$\pm$  0.1& 0.84 \\
       &  5.0$\pm$  0.0&  0.6$\pm$  0.0& 0.41&  5.9$\pm$  0.4& -0.8$\pm$  0.0& 0.59 \\
\hline
VCC1036&  2.9$\pm$  0.3&  0.3$\pm$  0.1& 0.71& 10.7$\pm$  4.7& -1.4$\pm$  0.3& 0.29 \\
       &  2.4$\pm$  0.3&  0.1$\pm$  0.1& 0.94& 18.0$\pm$ 14.7& -2.3$\pm$  2.9& 0.06 \\
       &  2.9$\pm$  0.1&  0.1$\pm$  0.0& 0.61& 17.2$\pm$  1.0& -1.7$\pm$  0.1& 0.39 \\
\hline
VCC1087&  4.9$\pm$  5.1&  0.7$\pm$  0.5& 0.16&  5.8$\pm$  1.4& -0.4$\pm$  0.2& 0.84 \\
       &  4.8$\pm$  1.7&  0.1$\pm$  0.2& 0.83&  8.3$\pm$  4.2& -2.3$\pm$  0.0& 0.17 \\
       &  0.1$\pm$  0.0&  0.0$\pm$  0.3& 0.01&  6.8$\pm$  0.6& -0.4$\pm$  0.0& 0.99 \\
\hline
VCC1261&  1.7$\pm$  0.4&  0.3$\pm$  0.1& 0.32& 12.0$\pm$  0.0& -1.1$\pm$  0.2& 0.68 \\
       &  2.4$\pm$  0.3&  0.0$\pm$  0.2& 0.74& 12.0$\pm$  0.0& -1.9$\pm$  0.9& 0.26 \\
       &  0.9$\pm$  0.1&  0.4$\pm$  0.1& 0.09& 12.0$\pm$  0.0& -0.8$\pm$  0.0& 0.91 \\
\hline
VCC1431&  5.0$\pm$  1.8& -1.0$\pm$  0.0& 0.14& 10.4$\pm$  2.4& -0.4$\pm$  0.0& 0.86 \\
       &  0.3$\pm$  0.7& -0.3$\pm$  2.5& 0.00& 13.8$\pm$  7.4& -0.6$\pm$  0.1& 1.00 \\
       &  0.1$\pm$  0.0& -0.1$\pm$  0.1& 0.00& 12.8$\pm$  1.0& -0.6$\pm$  0.0& 1.00 \\
\hline
VCC1861&  1.3$\pm$  0.6&  0.5$\pm$  0.3& 0.08&  7.2$\pm$  2.6& -0.4$\pm$  0.1& 0.92 \\
       &  3.6$\pm$  3.9&  0.7$\pm$  0.0& 0.46&  5.0$\pm$  0.0& -0.7$\pm$  0.4& 0.54 \\
       &  5.0$\pm$  0.0& -0.1$\pm$  0.1& 0.81&  6.1$\pm$  0.7& -1.8$\pm$  0.2& 0.19 \\
\hline
VCC2048&  4.3$\pm$  0.6& -0.1$\pm$  0.1& 0.70& 12.0$\pm$  0.0& -1.7$\pm$  0.1& 0.30 \\
       &  2.7$\pm$  0.4&  0.1$\pm$  0.2& 0.65& 12.0$\pm$  0.0& -1.7$\pm$  0.7& 0.35 \\
       &  1.1$\pm$  0.0&  0.1$\pm$  0.0& 0.11& 12.0$\pm$  0.0& -1.0$\pm$  0.0& 0.89 \\
\hline

\end{tabular}
\label{sdss-sauron-comparison-table} 
\end{table*}

\section{Methods}

To derive star-formation histories of our galaxies we use the full-spectrum fitting package ULySS of \cite{koleva:2009}. In ULySS, an observed spectrum is fitted against a model expressed as a linear combination of non-linear components (a function of age, [Fe/H], and [Mg/Fe]), returning the spectrum of a single stellar population), optionally convolved with a line-of-sight velocity distribution (LOSVD) and multiplied by a polynomial function, meant to absorb errors in, e.g. flux calibration (for details on ULySS see \cite{koleva:2009} sections 2 and 4). As reference models we use the extended MILES models (\citealt{vazdekis:2010}, \citealt{vazdekis:2012}) and Pegase.HR \citep{leborgne:2004}.

We first compared star formation histories extracted from the archival SDSS and our SAURON data. To this end, for each SAURON galaxy in common with SDSS spectroscopic data (10 objects) we extracted spectra collapsed within an aperture with a 3-arcsec diameter (see Figure~\ref{example-spectrum} for an example). We then compared the recovered star-formation histories (SFHs) coming from the following data sets (Figure~\ref{sfh-comparison}):
\begin{enumerate}
\item original SDSS,
\item SDSS restricted to SAURON wavelength range,
\item SAURON.
\end{enumerate}
The idea behind the approach was to test how/if restricting available wavelength range influences the accuracy of recovered SFHs and whether or not it is possible to compensate for a limited wavelength range with high S/N.

To produce the star formation  histories we use 2-SSP fits. We have found this to be the best decomposition given our data constraints. 1-SSP fits can be viewed as a first-order approach to the true SFHs, but going one step further with 2-SSP decompositions we are in the position to address the science questions that have driven our analysis. To determine the best strategy we first produced 20-SSP fits where individual population parameters were not allowed to change. From the non-zero weighs of these 20 SSPs we concluded that we can reconstruct the history in two episodes, except for NGC\,3073, for which we found that applying a 3-burst decomposition provided the most reliable recovered SF parameters. 

Then, using the MILES models \citep{vazdekis:2012} we looked for a two-burst decomposition in the age ranges [100, 5000]\,Myr and [5000,17000]\,Myr. We found that many of the galaxies hit the age limit in particular the 100 Myr and the maxiumum Z of 0.2 dex. We therefore tried to use Pegase.hr/ELODIE3.1 (\citealt{leborgne:2004}, \citealt{prugniel:2001}) which goes to 10\,Myr and metallicities of 0.69 (due to the different set of isochrones)\footnote{as both the libraries are empirical, a full coverage of parameter space is not possible, therefore the quoted limits are reached through extrapolations performed with the use of theoretical libraries.}. This was more successful as some of the galaxies found their best young population fits around 50\,Myr so using ELODIE we made sure that the model parameters were fully enclosed within the model grid.

In our subsequent reconstructions the ages of old populations were either left free or were fixed to 12\,Gyr. Young populations' ages were left free in both cases. The metallicities were in each case left free within the limits of the models.
When fitting, we have included SSPs together with H$\beta$, [O\,III\,4959] and [O\,III\,5007] lines in emission. We note, however, that significant emission was only detected in one of our field objects, NGC\,3073. See also \cite{rys:2013} for more details.

For those of the galaxies for which the program turned out to have difficulty finding an old population (VCC\,0523, VCC\,1087, VCC\,1261, VCC\,1861, VCC\,2048, and ID\,0650), we imposed more constraints by fixing the age of the old burst to 12 Gyr. The decision was motivated by the fact that old populations have so far been found in all studied dwarfs, not only in a comparable mass range (e.g. \citealt{koleva:2009}, \citeyear{koleva:2011}) but also in  in all dwarf early types and the vast majority (possibly all) late- and transition-type dwarfs of the Local Group (\citealt{tolstoy:2009}). The difficulty, thus, most likely stems from the strong presence of younger populations, rather than a genuine lack of old stars.

The mass weight contained in each annulus comes from the fraction of each component used to obtain the best fit. Each of the SSPs is multiplied by this fraction and later the SSPs are summed to produce the best matching template. In Figure~\ref{sfh-profiles-weights} we show the relative contribution of each component, i.e. the weight of each SSP is normalized to the total weight.

As a consistency check, we have also compared our ULySS-based light-weighted ages and metallicites with those coming from the line-strength analysis which will be presented in our forthcoming paper (Ry\'s et al., in prep.). The comparison plots and their description are given in Appendix~\ref{app-a}. Overall, the values derived using the two methods agree to within the errors. 

The SFHs for all galaxies and radial bins, including errors on the recovered parameters coming from Monte-Carlo simulations, are tabulated in Appendix B.

\section{Results}

The goal of the stellar population analysis is to gain insight into the star-formation and assembly history. Our approach is twofold. Here we analyze the radial  properties as well as integrated (central) population parameters of our galaxies, in search of commonalities and/or differences within the sample. In the subsequent paper (Ry\'s et al., in prep.) we will take these results and, together with the results from line-strength analysis, will juxtapose them with those for more massive galaxies to look for potential (dis)similarities between the various galaxy classes. 

\subsection{SAURON vs. SDSS star formation histories}

Star formation histories were until recently extracted only in the cases where significantly longer wavelength ($\lambda$) ranges were available. This is generally required given the inherent degeneracies in stellar population parameters where ages and metallicities cannot be easily separated. One thus aims to include as broad a range of features (i.e. have a long $\lambda$ coverage) as possible to minimize these effects. With SAURON the available wavelength range (4760-5300\,\AA) is rather limited, however, high S/N values can provide a leverage against the $\lambda$ range in the context of SFH recovery. This was recently tested for giant early-type galaxies from the ATLAS3D sample by \cite{mcdermid:2015}. Their results show that there is no systematic bias between the SDSS and SAURON extracted SFHs (see their Section 2.6 for details.)

To see whether the results would hold also for our low-mass galaxies, we have extracted SFHs for the SDSS data, which provide central 3-arcsec spectroscopic information for a number of galaxies as a part of the survey.  SDSS presented itself as our natural choice, given that its spectroscopic sample had 10 objects in common with ours and the available wavelength range is extended (3800-9200\,\AA\footnote{https://www.sdss3.org/dr9/spectro/spectro\_basics.php}). At the same time SDSS data have significantly lower S/N ratios given predominantly much shorter exposure times (between 2400\,s and 5400\,s). The results of the analysis are shown in Figure~\ref{sfh-comparison} and tabulated in Table~\ref{sdss-sauron-comparison-table}. The three datasets were compared with one another and for the majority of objects the recovered ages, metallicities and mass weights (i.e the relative amounts of stellar masses contained in old and young populations) agree to within the errors. Outlying values belong to different datasets for different galaxies, that is, there are no systematic trends present. We note however, that for the second dataset  -- SDSS with the SAURON $\lambda$ range -- the values tend to deviate from the other more often, suggesting that in this case low S/N \textit{together with} limited $\lambda$ range increase the uncertainties. 

We can see that, in general, the recovery of metallicity for the younger component is quite good. This is expected as the age/metallicity degeneracy is more severe for old populations. The strongest outlier (middle left panel) is VCC\,1431 for which we find nearly null fractions of the younger population, which is later confirmed in the full radial profiles analysis in the subsequent sections.

We conclude that  agreement  between the datasets is reasonably good,  most of the existing discrepancies are caused by random errors and that our results presented in the subsequent sections are robust. Nevertheless, we note that the main focus of the exercise was to determine the suitability of the rather limited SAURON range, rather than an absolute comparison between the the data sets. 

\subsection{Star-formation histories}

Figure~\ref{sfh-profiles} shows 2-SSP star-formation histories for our sample, where the ages and metallicities of both young and old components are shown for each galaxy (columns 1 and 2) together with the mass and light weights of both populations (columns 3 and 4).

Among the galaxies for which we did not fix the age of the old population to 12\,Gyr, for two of the objects we found non-negligible contributions from intermediate-aged components, with ages between [3,10]\,Gyr. For the remaining four (VCC\,0308, VCC\,1431, NGC\,3073, and ID\,0918) we found ages in the range of $\gtrsim$\,10-13\,Gyr.

The old components of VCC\,1431 and VCC\,0308 have the maximum allowed ages. VCC\,0308 is our only ``blue center'' galaxy in the sample, i.e. it is an object whose photometry indicates recent central star formation episodes and indeed the age of the young population is as low as $<$\,1\,Gyr in the center. VCC\,1431 is an interesting case since here the recovered age of the younger population  practically coincides with that of the old one. No significant SF activity seems to have taken place there at least in the last 10\,Gyr in the center and 12-13\,Gyr in the outskirts.

In eight objects the old population dominates the mass at all radii. Interestingly, in the case of two galaxies (VCC\,0523 and VCC\,1261) the relative contribution of younger population increases with radius. In another two (VCC\,1861 and  less strongly VCC\,2048) we see a young population ''bump'' at around 0.2-0.4\,$R_{e}$, which subsequently goes to zero. The remaining weight profiles are flat to within the errors. 

There are positive age gradients present in the younger populations of seven of our galaxies, with the remaining five showing the overall gradient values consistent with zero. In the case of VCC\,0929, belonging to the latter group, there is an increase in the young population age at around 0.1\,$R_e$ and the profile flattens out for larger radii. The population's weights are, however, nearly zero, so we do not interpret the above feature as significant. A similar, though shallower, feature can be seen in the profile of VCC\,1861. Noting that the young population weights drop to zero for larger radii we can conclude that the central 0.3\,$R_e$ shows a young population presence with a positive age gradient.

We see a slight age gradient in the older population of VCC\,1036 (of an intermediate age in this case) and a compound age gradient in ID\,0918, where a clear positive radial age trend is broken at $\sim$\,1.2\,$R_e$ for both populations, with the ages being much lower for larger radii but again seeming to pick up a positive trend there as well. All the other gradients are nearly or fully consistent with zero. We note that this is not an indication of the lack of gradients but rather a result of the inherent difficulties in model resolution at those higher ages. 

Metallicities of young populations are higher than or equal to those of old populations, with the former on average around the solar value in the centers and declining outwards or remaining constant, and the latter approximately in the range (0,-2) and showing a similar outward trend.

One third of our objects show the presence of a metallicity gradient in their older populations. These are VCC\,0523, VCC\,1087 (where the ages themselves are fixed to 12\,Gyr), VCC\,1036 and VCC\,1261 (where the ages are intermediate, i.e, up to 10\,Gyr). For the remaining objects the gradients are nearly or fully consistent with zero.

Young populations show a larger spread in metallicity gradient values. Half of the sample shows negative gradients, four of which are strong, i.e. with $\nabla_Z$\,$>$\,0.5\,dex. Three objects have gradients consistent with zero and the remaining three have positive ones. To the last group belongs ID\,0918 where the trend is actually composed of two flat profiles with a break at around 1.2\,$R_e$: the outer profile is higher by $\sim$0.8\,dex. A similar but less pronounced trend is seen in VCC\,2048, this time with the difference being around 0.4\,dex and the break radius at $\sim$0.6\,$R_e$. Here also the weights of the younger component fall to nearly zero above the quoted radius. So it is more correct to interpret the profiles as having a young component that is restricted to the inner 0.5\,$R_e$. NGC\,3073's positive metallicity trend may be related to the galaxy being in the vicinity and under the tidal influence of the larger NGC\,3079, but this would require some mixing of stellar material coming from the large neighbor.

\begin{figure*}
\centering
\hspace{1.0cm}Age \hspace{3.2cm} Metallicity \hspace{3.4cm} Mass weight \hspace{2.9cm} Light weight\\
\vspace{0.3cm}

\includegraphics[width=0.48\columnwidth]{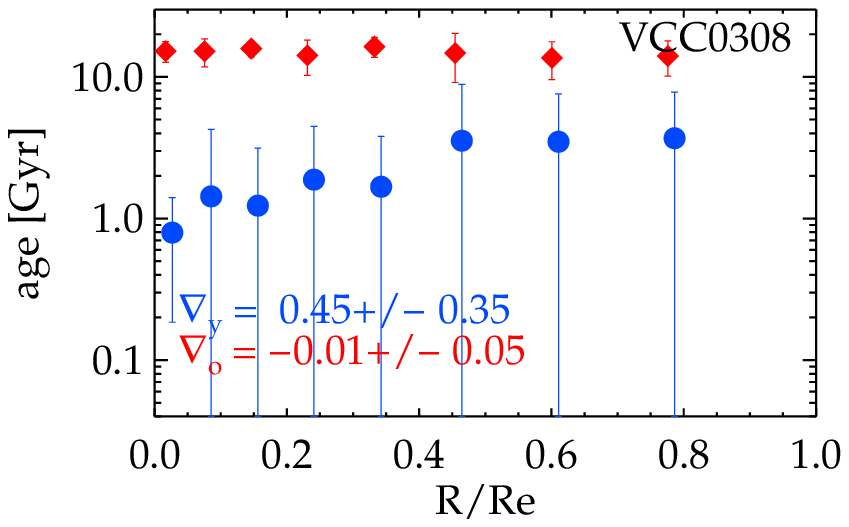}\hspace{0.2cm}
\includegraphics[width=0.48\columnwidth]{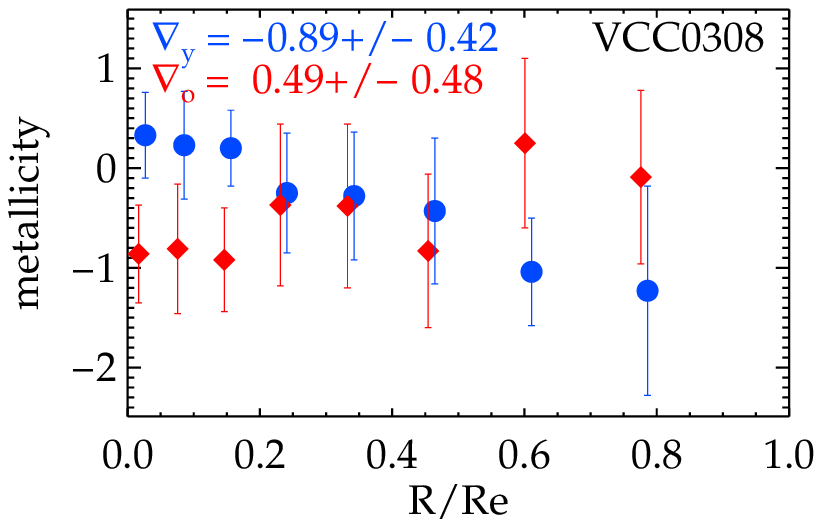}\hspace{0.8cm}
\includegraphics[width=0.48\columnwidth]{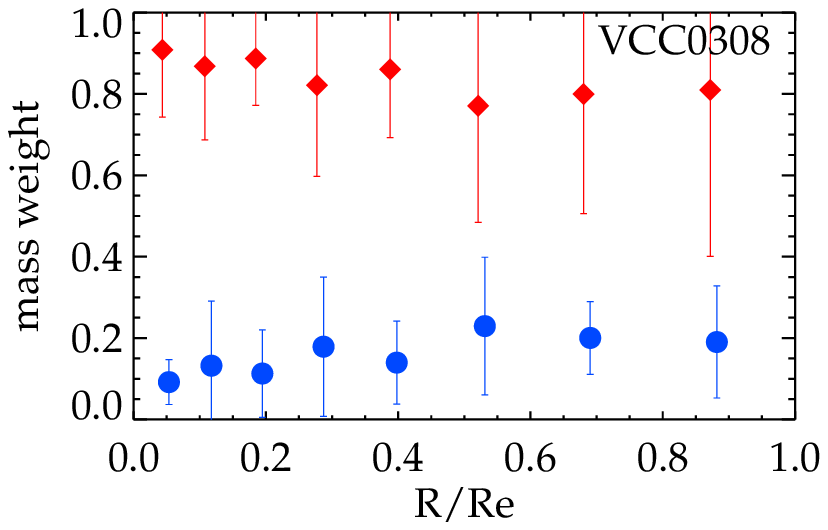}\hspace{0.2cm}
\includegraphics[width=0.48\columnwidth]{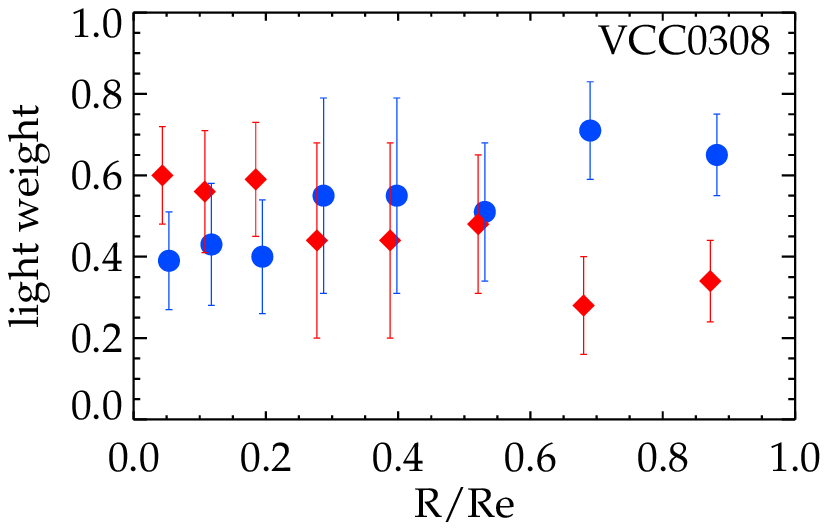}
\vspace{0.8cm}

\includegraphics[width=0.48\columnwidth]{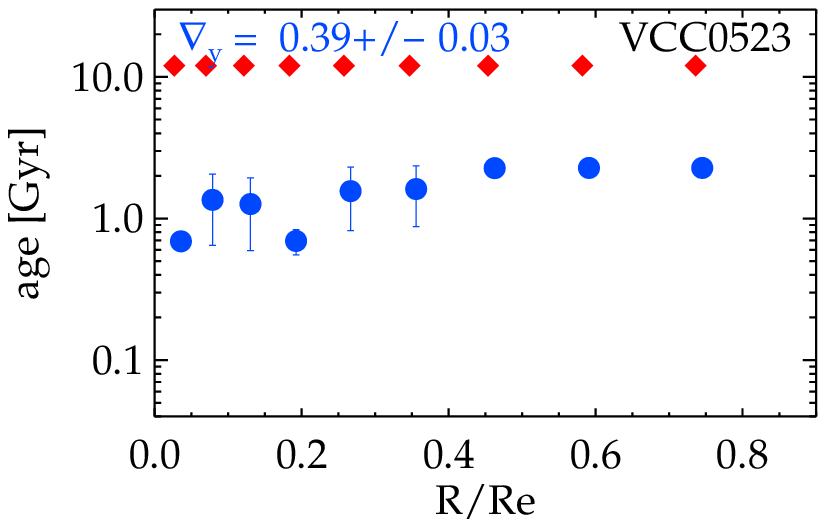}\hspace{0.2cm}
\includegraphics[width=0.48\columnwidth]{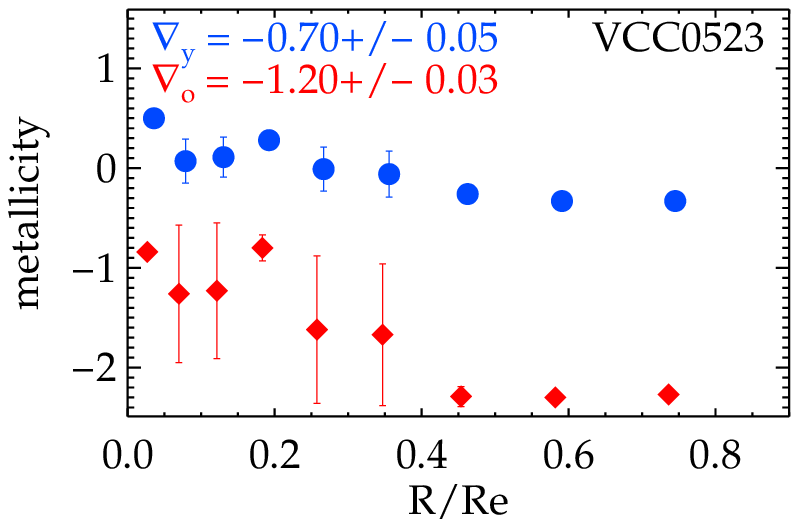}\hspace{0.8cm}
\includegraphics[width=0.48\columnwidth]{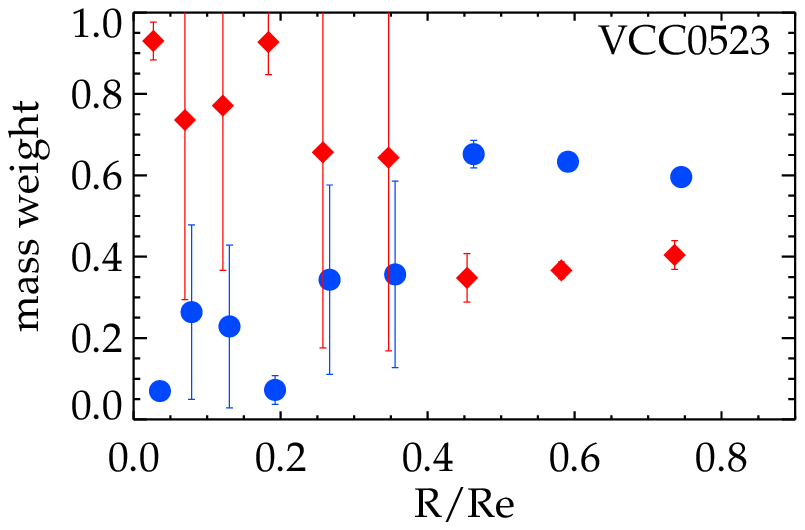}\hspace{0.2cm}
\includegraphics[width=0.48\columnwidth]{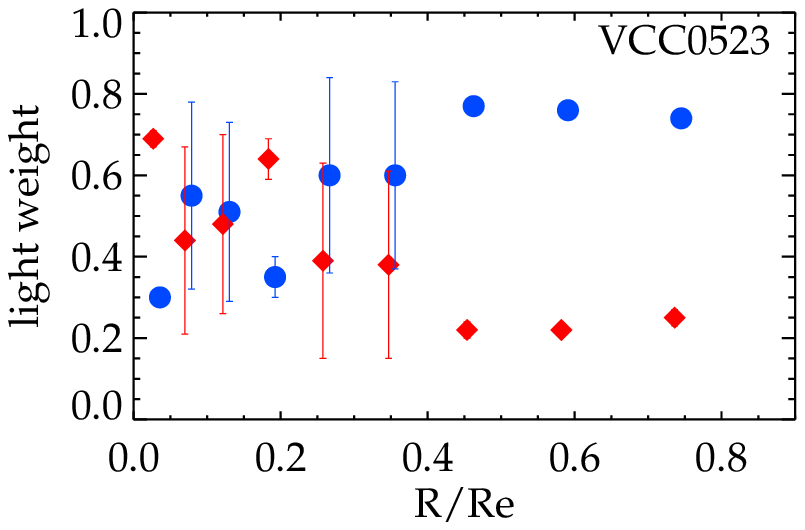}
\vspace{0.8cm}

\includegraphics[width=0.48\columnwidth]{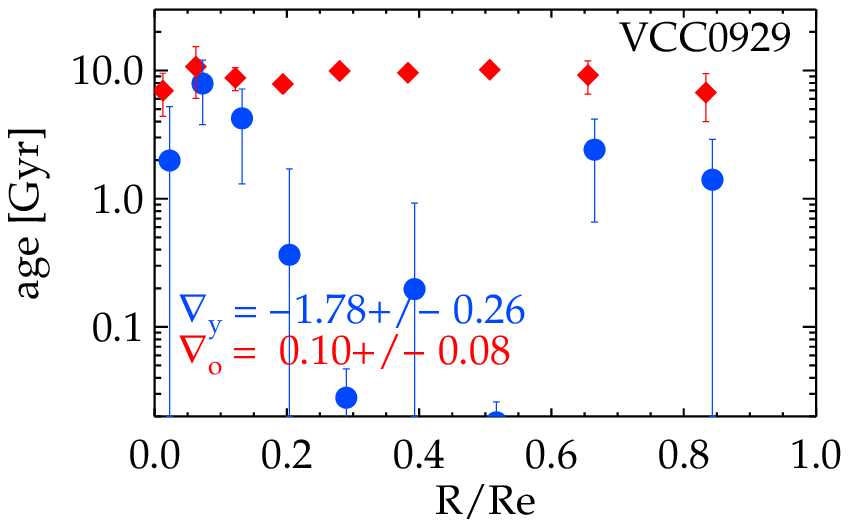}\hspace{0.2cm}
\includegraphics[width=0.48\columnwidth]{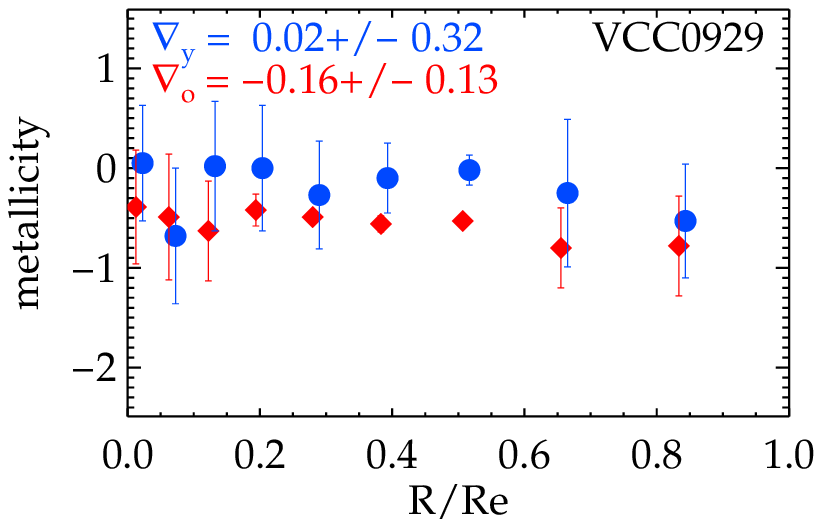}\hspace{0.8cm}
\includegraphics[width=0.48\columnwidth]{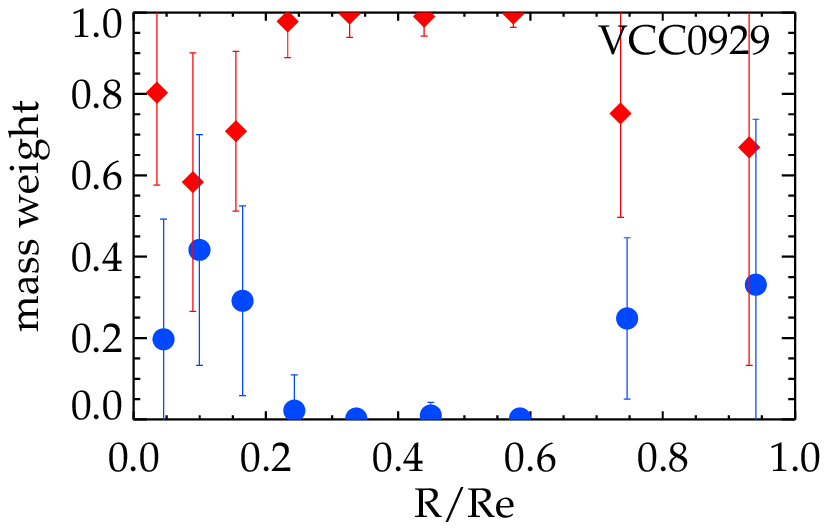}\hspace{0.2cm}
\includegraphics[width=0.48\columnwidth]{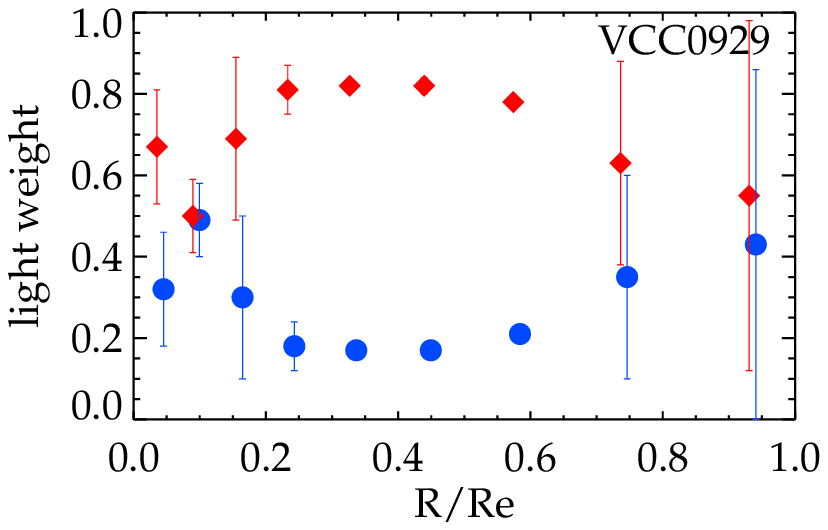}
\vspace{0.8cm}

\includegraphics[width=0.48\columnwidth]{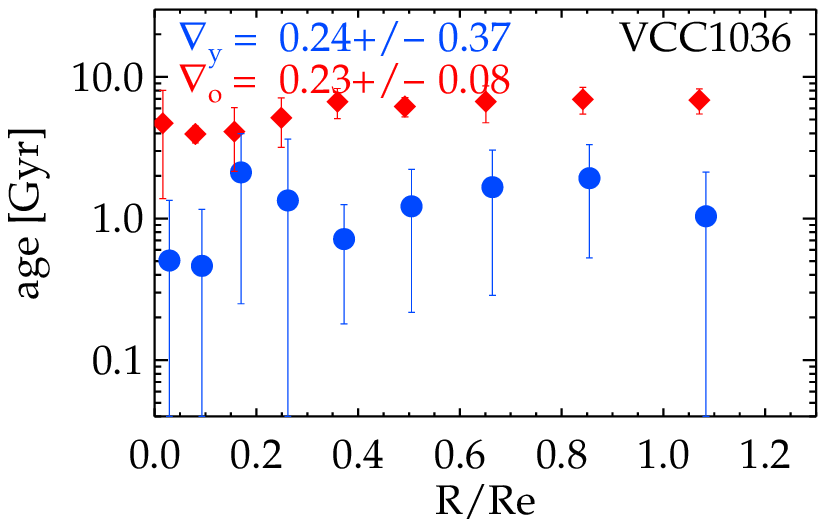}\hspace{0.2cm}
\includegraphics[width=0.48\columnwidth]{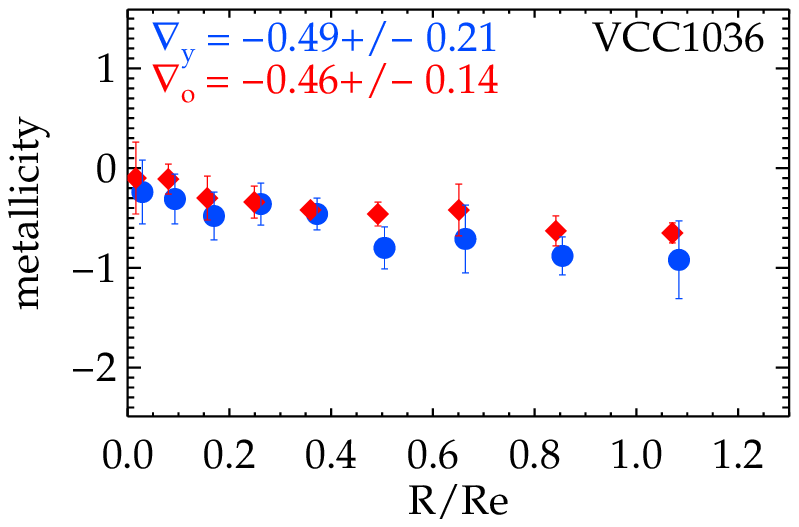}\hspace{0.8cm}
\includegraphics[width=0.48\columnwidth]{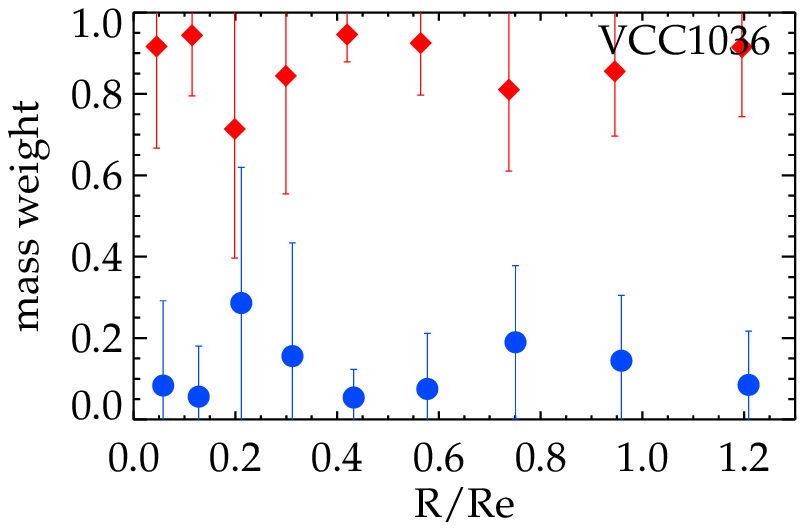}\hspace{0.2cm}
\includegraphics[width=0.48\columnwidth]{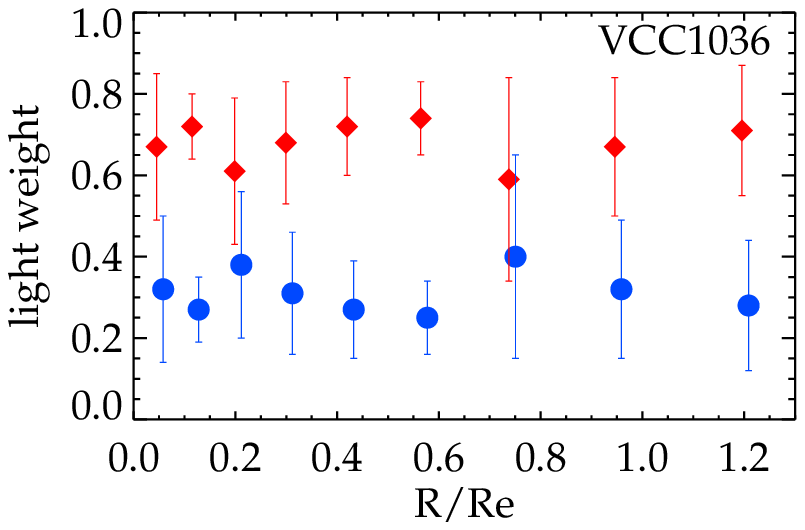}
\vspace{0.8cm}

\includegraphics[width=0.48\columnwidth]{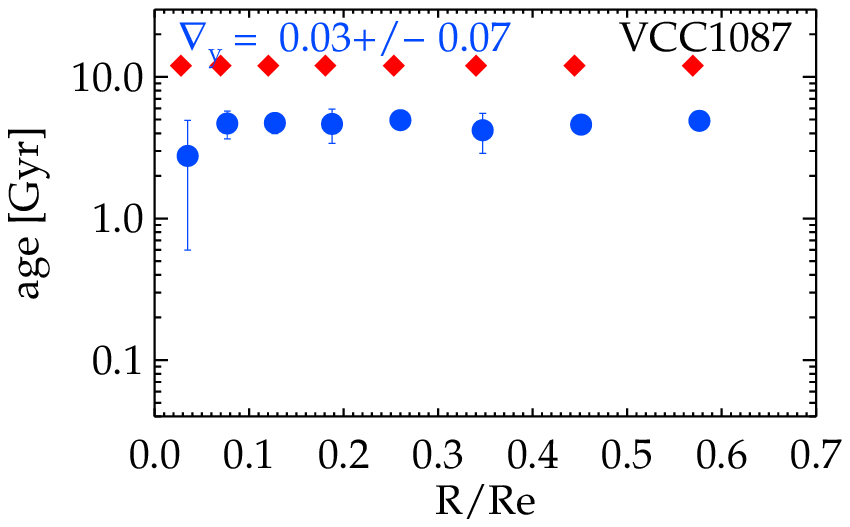}\hspace{0.2cm}
\includegraphics[width=0.48\columnwidth]{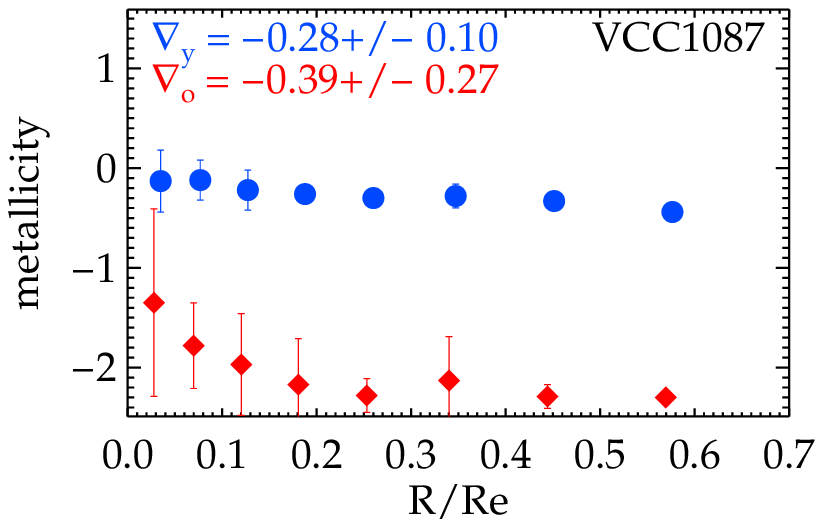}\hspace{0.8cm}
\includegraphics[width=0.48\columnwidth]{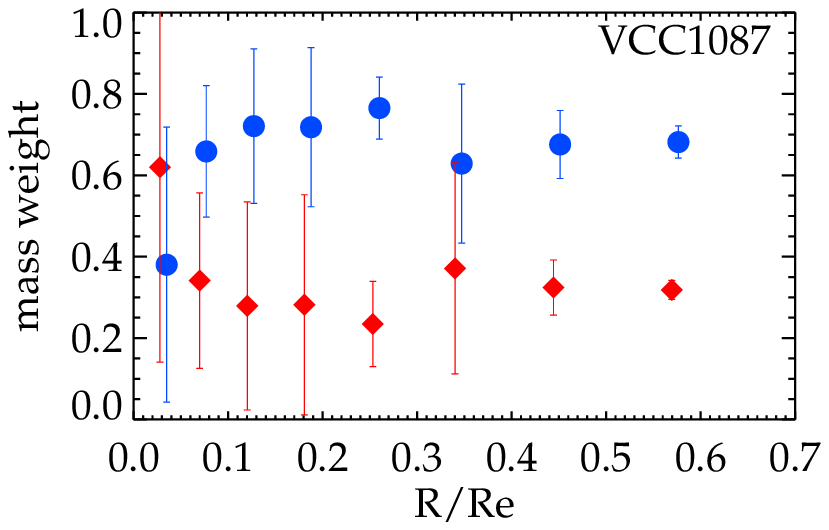}\hspace{0.2cm}
\includegraphics[width=0.48\columnwidth]{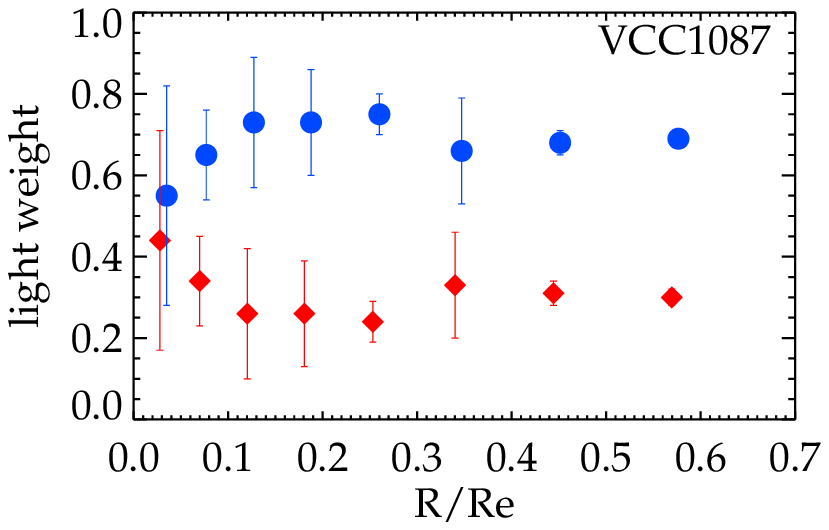}
\vspace{0.8cm}

\includegraphics[width=0.48\columnwidth]{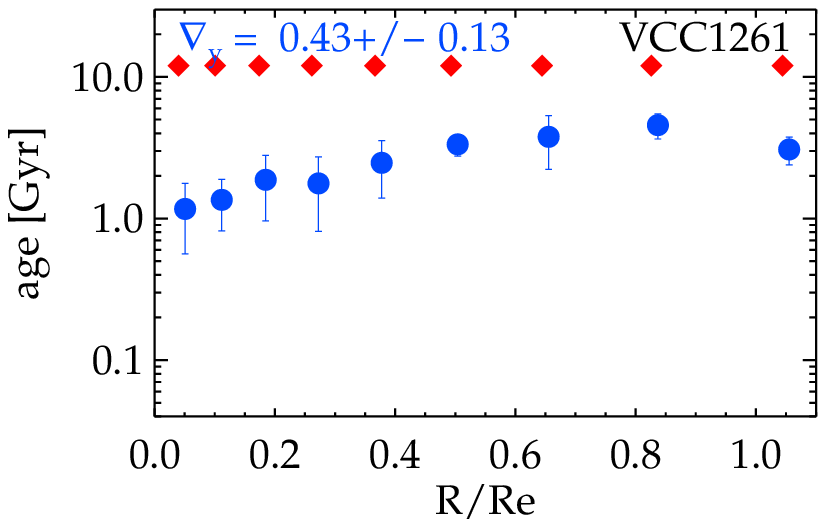}\hspace{0.2cm}
\includegraphics[width=0.48\columnwidth]{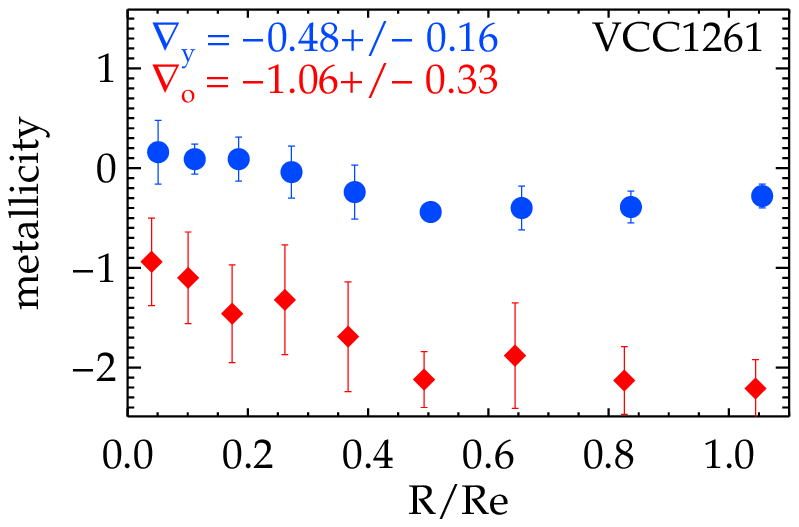}\hspace{0.8cm}
\includegraphics[width=0.48\columnwidth]{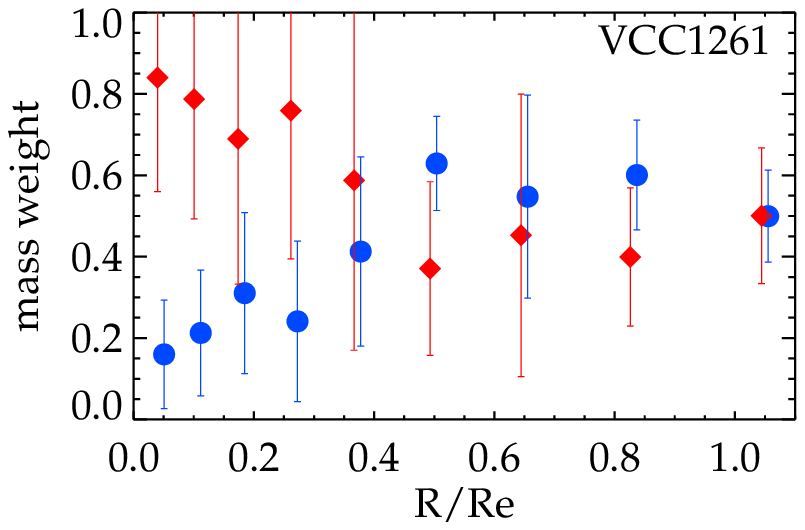}\hspace{0.2cm}
\includegraphics[width=0.48\columnwidth]{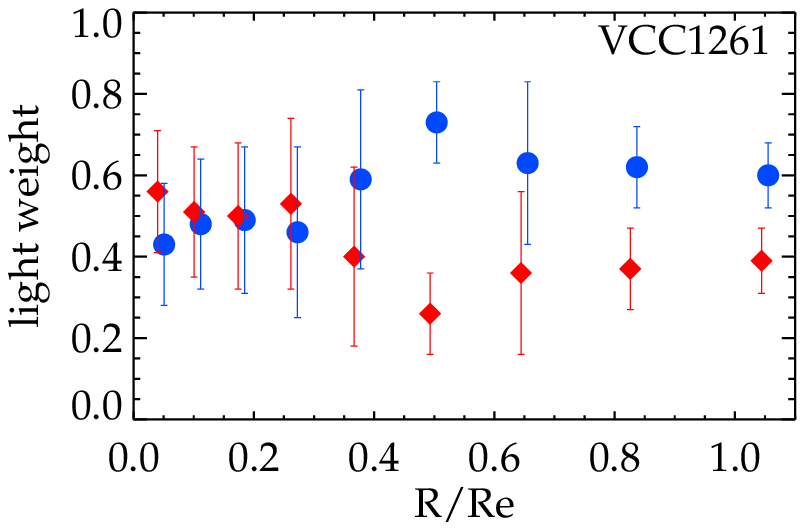}
\vspace{0.3cm}

\caption{Star formation histories extracted from our 2D data collapsed in annuli of increasing width. Younger populations are shown with blue circles and the older ones with red diamonds (for NGC\,3073 the intermediate population is shown with green triangles). The four columns show (respectively) age, metallicity (total metallicity in solar units), mass weight and light weight values, together with their errors, as a function of major axis effective radius. For VCC\,0308, VCC\,0929, VCC\,1036, VCC\,1431, NGC\,3073 and ID\,0918 we show fits for which no constraints were placed on the ages of the SSPs. For the remaining six galaxies the age of the older population was fixed to 12\,Gyr. The given weight values are relative, i.e. at any given radius they add up to 1.0. A small horizontal offset was added to the old population so as to make the visual appearance of the points and their errors clearer.}
\label{sfh-profiles}  
\end{figure*}

\addtocounter{figure}{-1}

\begin{figure*}
\centering
\hspace{1.0cm}Age \hspace{3.2cm} Metallicity \hspace{3.4cm} Mass weight \hspace{2.9cm} Light weight\\
\vspace{0.3cm}

\includegraphics[width=0.48\columnwidth]{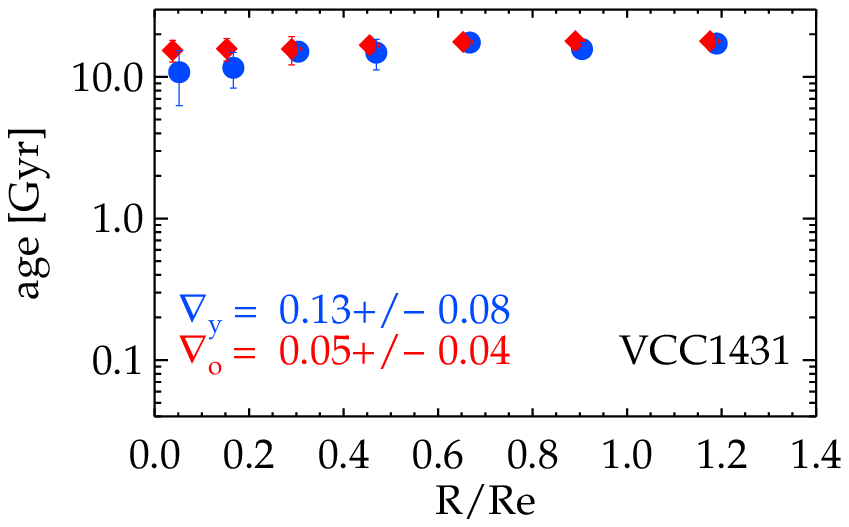}\hspace{0.2cm}
\includegraphics[width=0.48\columnwidth]{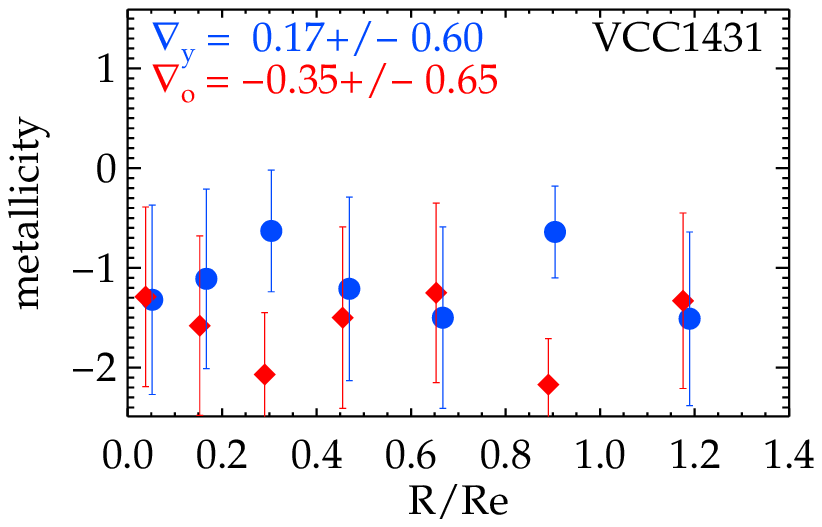}\hspace{0.8cm}
\includegraphics[width=0.48\columnwidth]{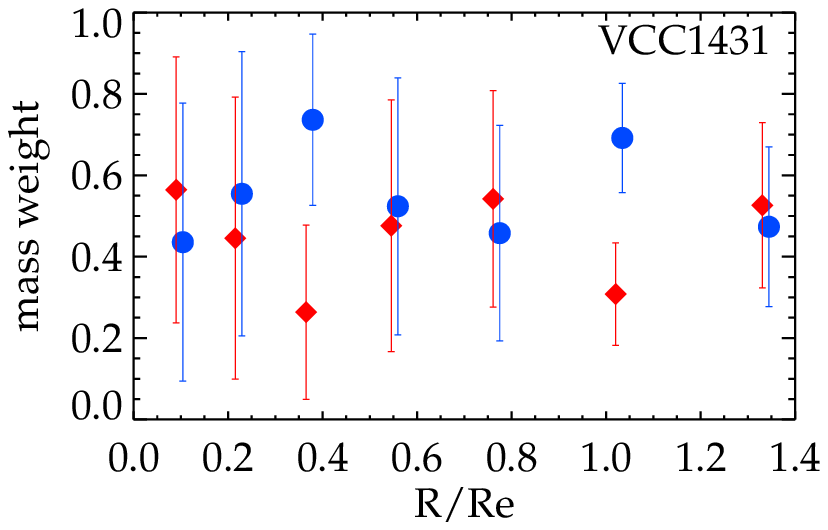}\hspace{0.2cm}
\includegraphics[width=0.48\columnwidth]{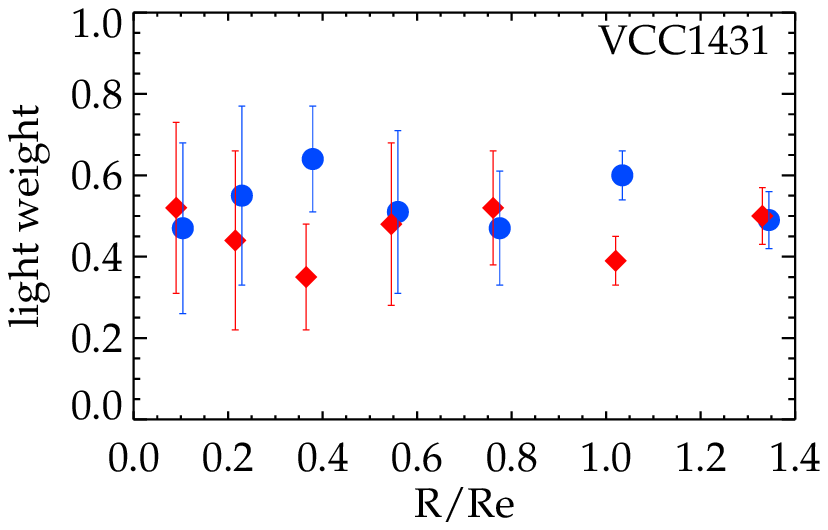}
\vspace{0.8cm}

\includegraphics[width=0.48\columnwidth]{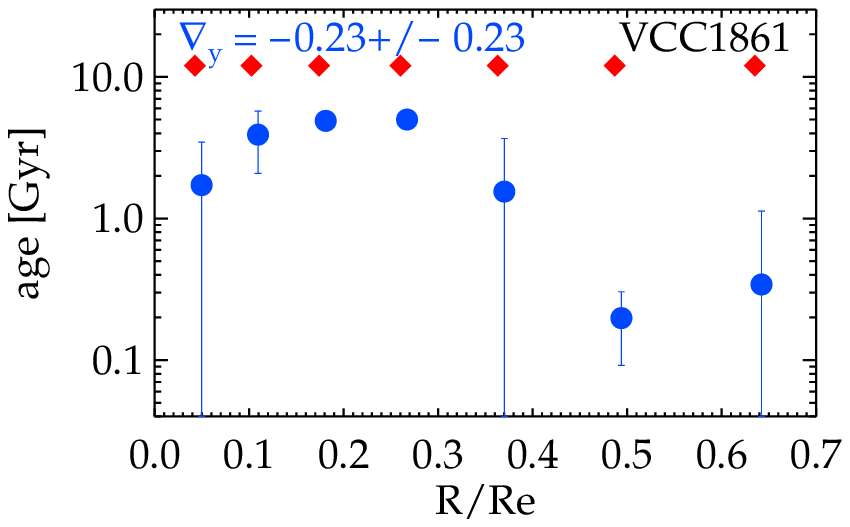}\hspace{0.2cm}
\includegraphics[width=0.48\columnwidth]{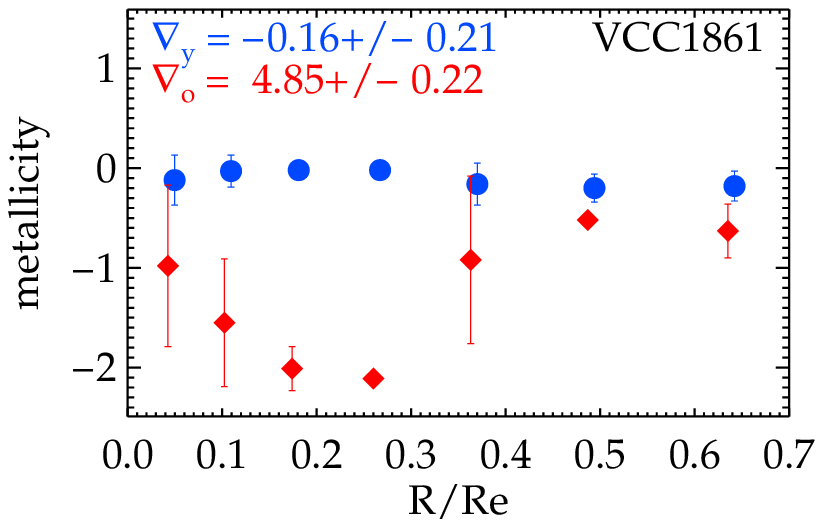}\hspace{0.8cm}
\includegraphics[width=0.48\columnwidth]{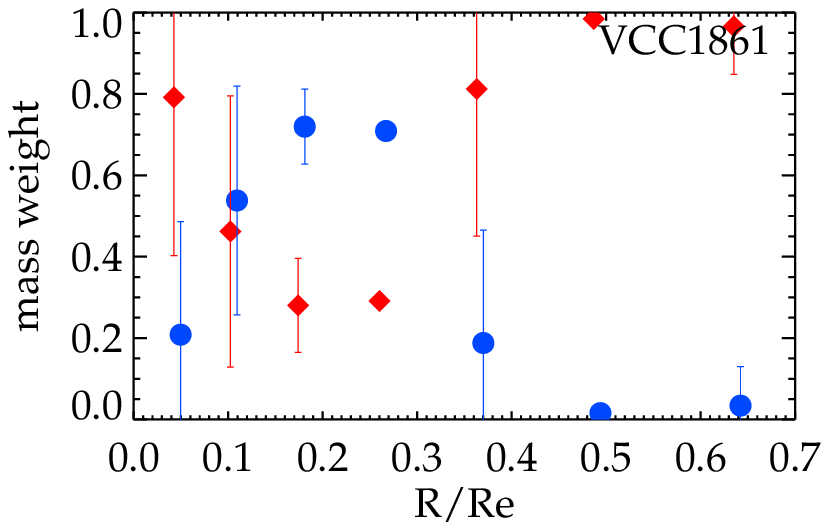}\hspace{0.2cm}
\includegraphics[width=0.48\columnwidth]{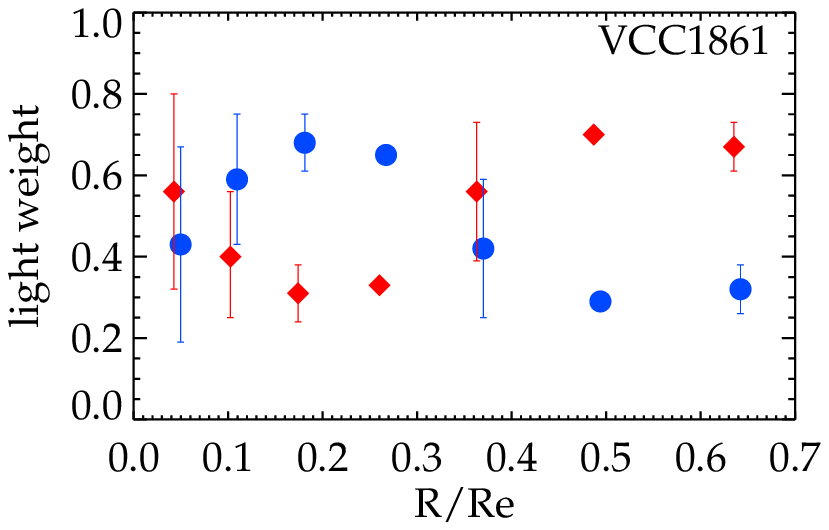}
\vspace{0.8cm}

\includegraphics[width=0.48\columnwidth]{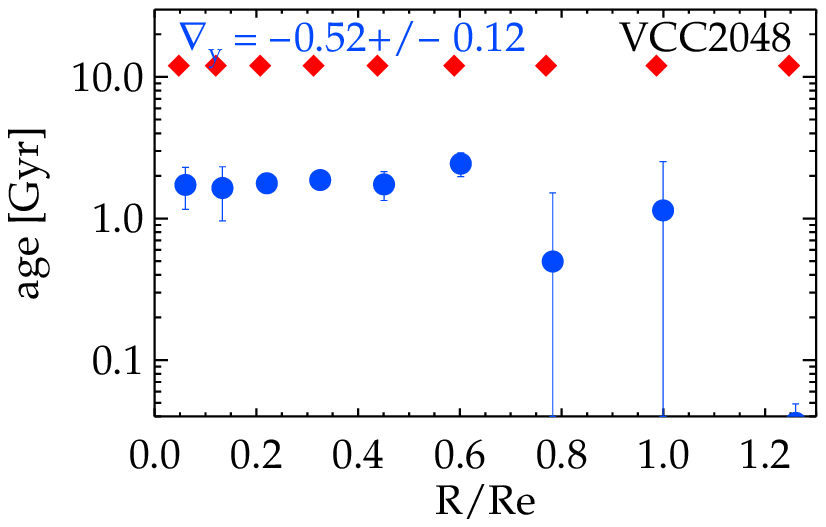}\hspace{0.2cm}
\includegraphics[width=0.48\columnwidth]{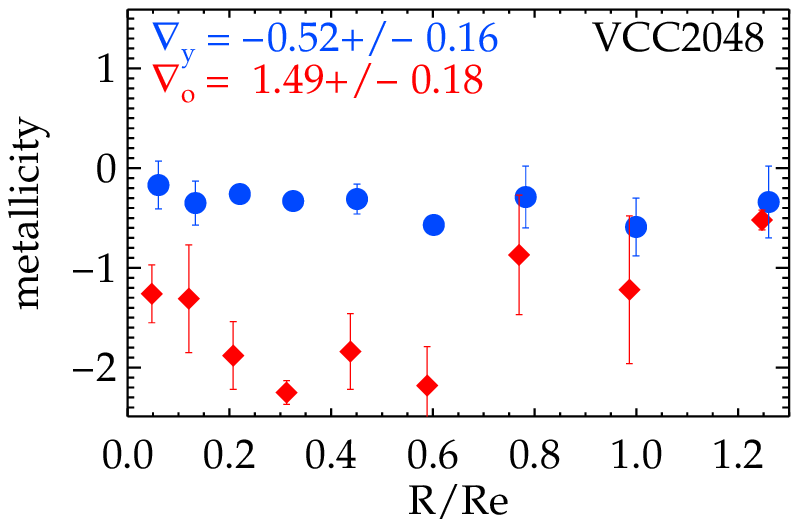}\hspace{0.8cm}
\includegraphics[width=0.48\columnwidth]{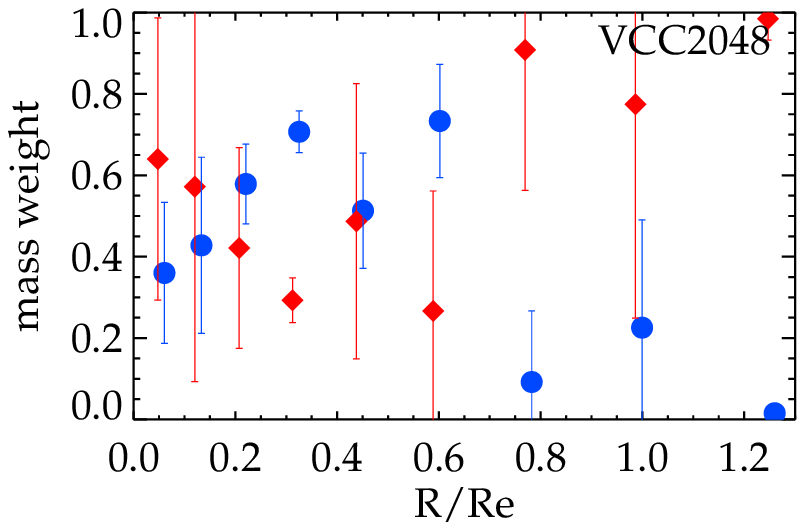}\hspace{0.2cm}
\includegraphics[width=0.48\columnwidth]{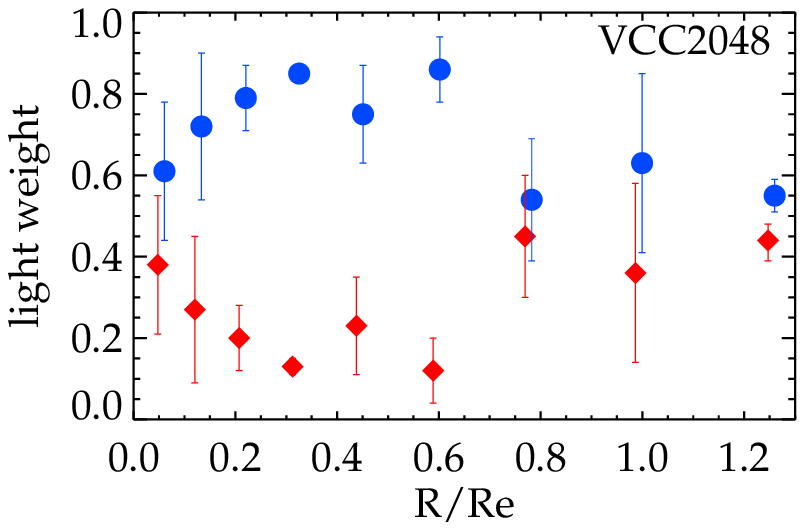}
\vspace{0.8cm}

\includegraphics[width=0.48\columnwidth]{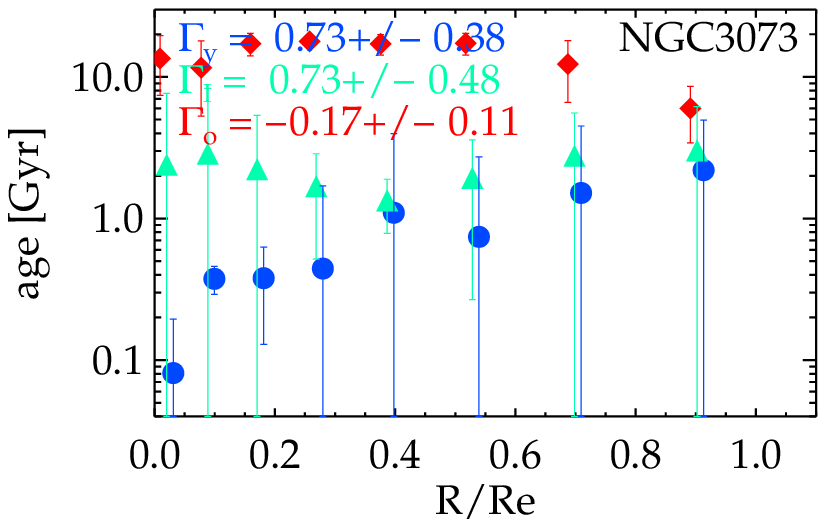}\hspace{0.2cm}
\includegraphics[width=0.48\columnwidth]{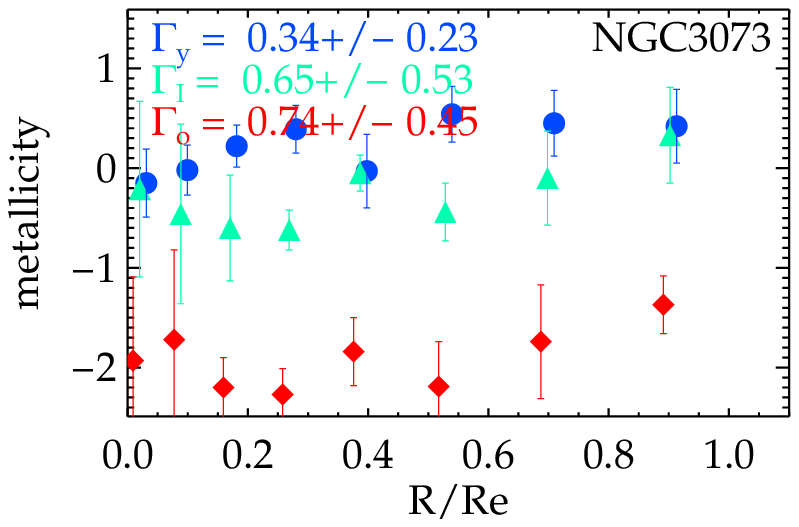}\hspace{0.8cm}
\includegraphics[width=0.48\columnwidth]{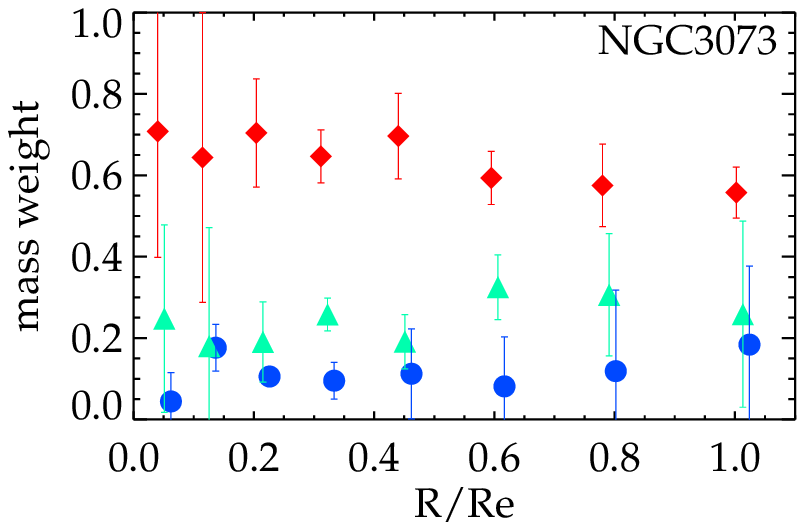}\hspace{0.2cm}
\includegraphics[width=0.48\columnwidth]{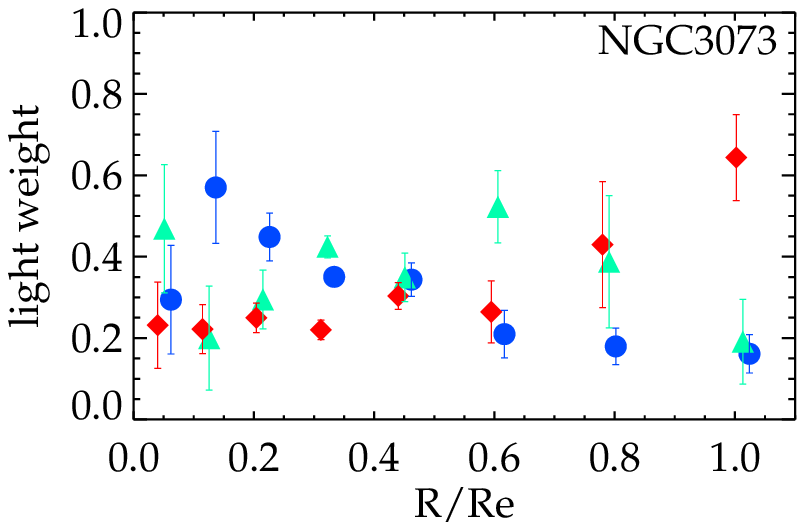}

\vspace{0.8cm}

\includegraphics[width=0.48\columnwidth]{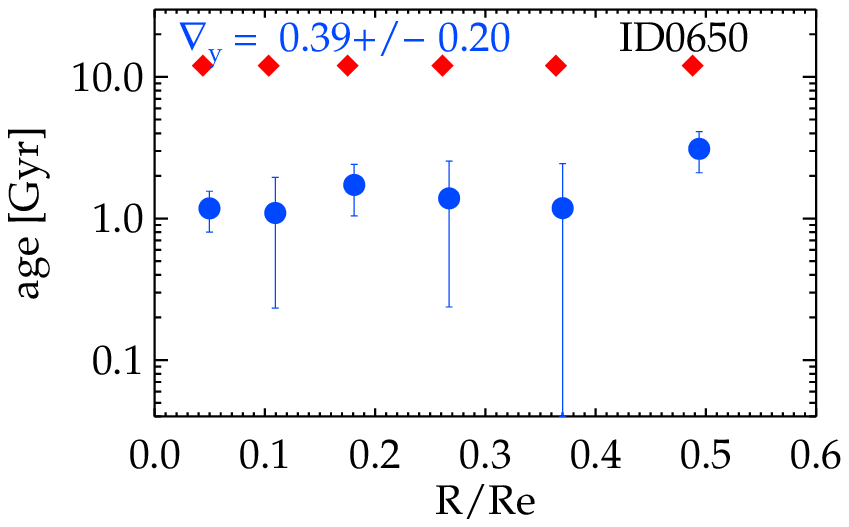}\hspace{0.2cm}
\includegraphics[width=0.48\columnwidth]{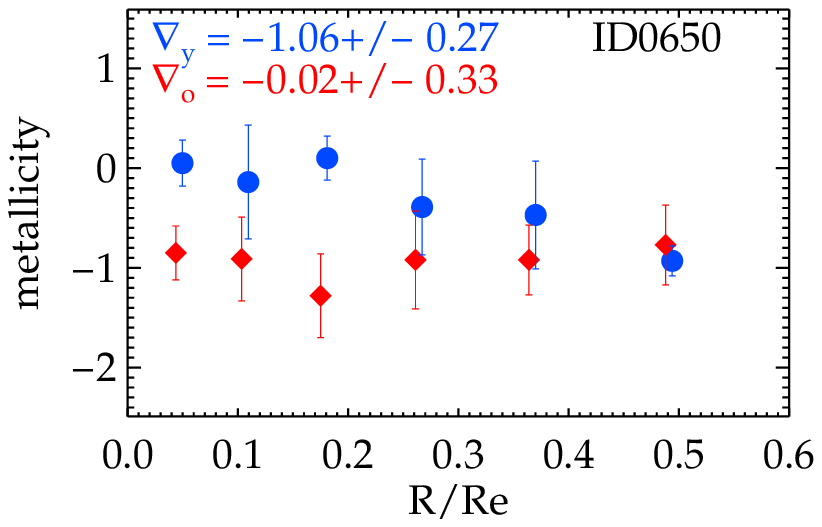}\hspace{0.8cm}
\includegraphics[width=0.48\columnwidth]{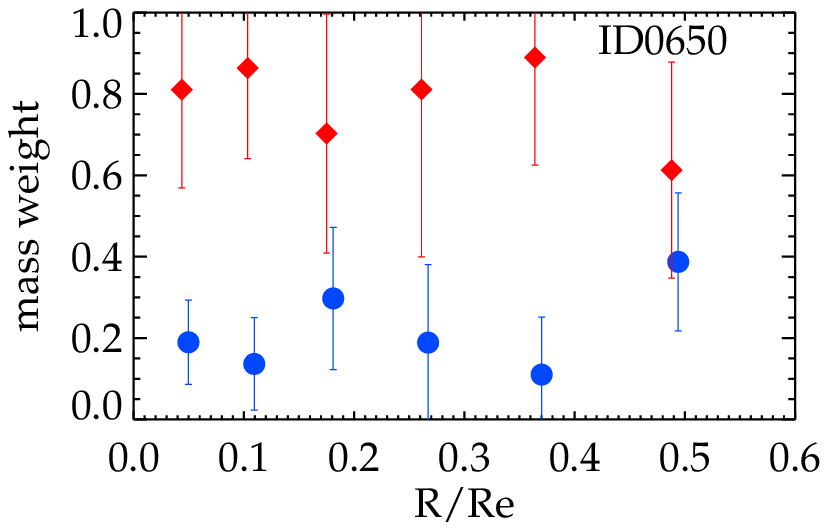}\hspace{0.2cm}
\includegraphics[width=0.48\columnwidth]{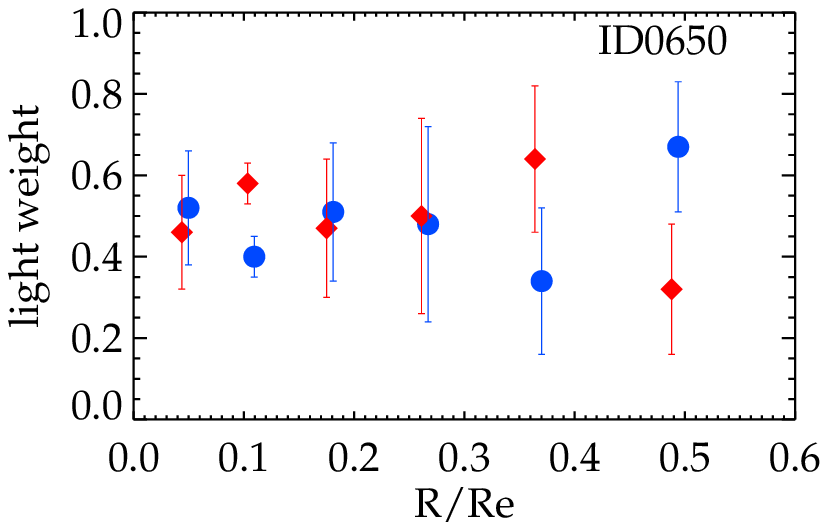}
\vspace{0.8cm}

\includegraphics[width=0.48\columnwidth]{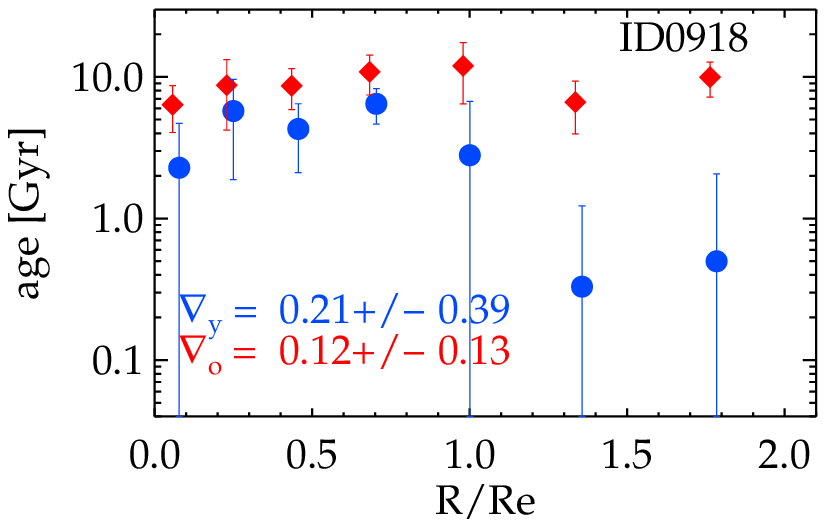}\hspace{0.2cm}
\includegraphics[width=0.48\columnwidth]{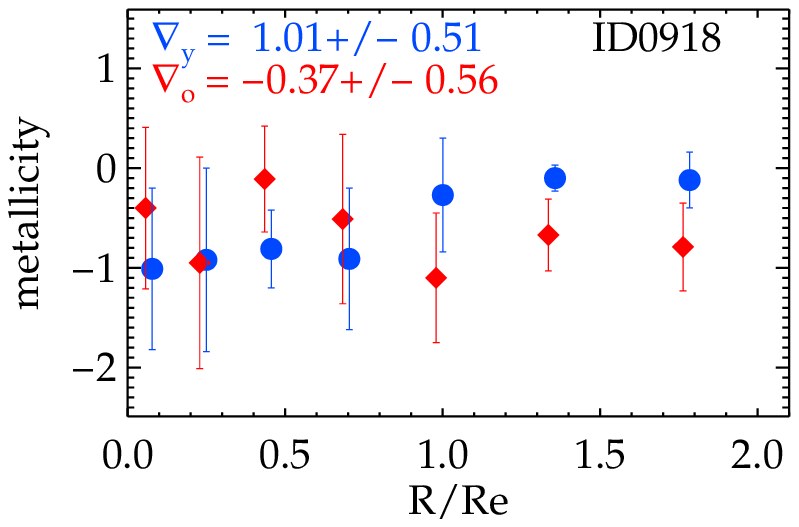}\hspace{0.8cm}
\includegraphics[width=0.48\columnwidth]{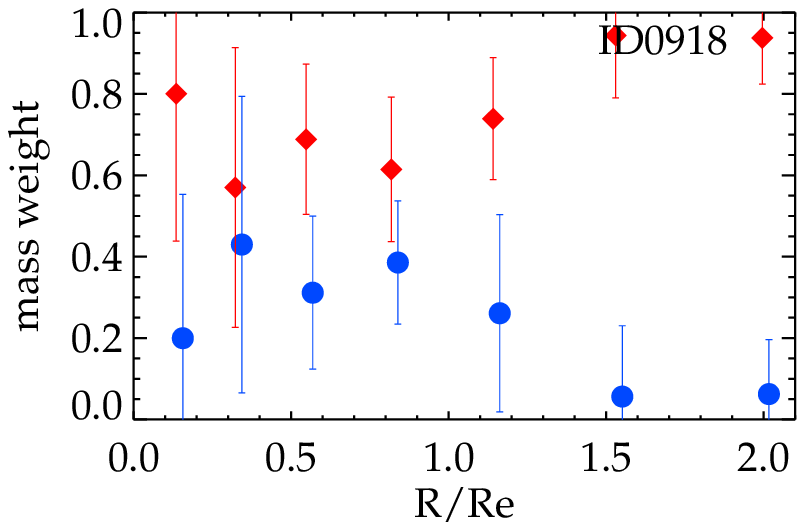}\hspace{0.2cm}
\includegraphics[width=0.48\columnwidth]{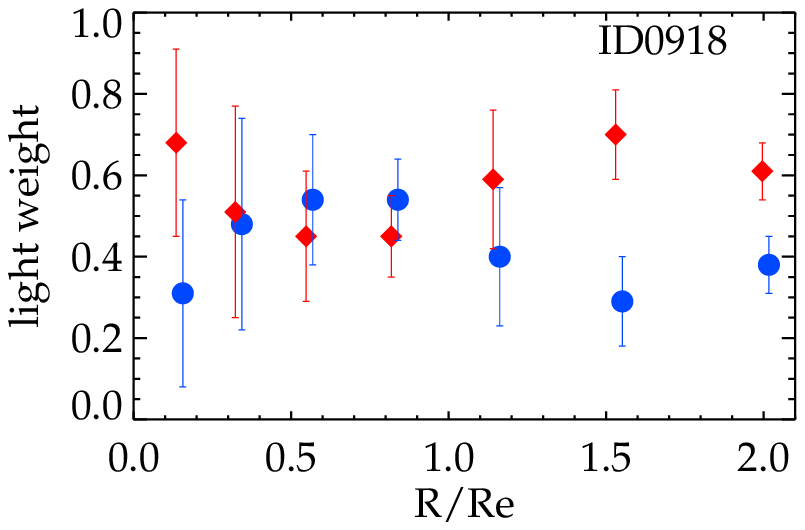}
\vspace{0.3cm}

\caption{Continued.}
\label{sfh-profiles-weights}  
\end{figure*}

\section{Discussion and Conclusions}

\subsection{Star formation histories}

First attempts at providing star formation histories of Virgo dEs are in \citeauthor{lisker:2006b} (\citeyear{lisker:2006b}b) where they used the 3-arcsec SDSS spectra for a sample of 16 dEs, 9 of which were classified as hosting a blue core. Not meant to be a fully-fledged SFH analysis, the results provided rough estimates on the age ranges of the studied objects. The study showed that all their galaxies had an underlying old population (with the age fixed at 5\,Gyr) in their centers, accounting for at least 90\% of the total mass, albeit contributing only a small fraction to the total galaxy light there. \cite{koleva:2009b} were the first to attempt a full SFH reconstruction, also spatially resolving their data into two bins (central and 1\,$R_e$ apertures). These authors confirmed the presence of old ($\gtrsim$\,10\,Gyr) populations in all but one of their galaxies. Interestingly, for about half of their sample they find constant/slowly declining SFHs, indicative of fairly continuous star formation activity from the early ages until recently. They concluded that the galaxies could not have been stripped of their gas early on in their evolution and that the ages of old populations are comparable with those of massive ETGs, with the difference being that dE SF efficiency is lower. This also holds true for lower mass systems in the Local Group (LG) where all studied dwarf types have shown evidence for the presence of older ages (see the review of \citealt{tolstoy:2009} and references therein).

Our results agree with those previous findings. We, too, find old or intermediate populations in all our objects, even more so, one of our galaxies seems to contain \textit{only} an old population with the age over $\gtrsim$\,10\,Gyr (see the following subsection). Interestingly, very similar results are found by \cite{hidalgo:2013} in their SFH results for four LG dwarfs (spheroidal and transitional type): the oldest populations are coeval across all probed radii, with the young components more prominent and younger in the inner regions (see their Figures~9 and~10). The mass difference between their and our sample is, naturally, significant, yet, the profile similarities are remarkable.

Seeing extended SFHs and/or multiple SF episodes in the majority of our objects further strengthens the conclusion of Paper II where we show that dEs have higher concentration and steeper rotation curves than their presumed late-type progenitors. We said there that it could be either due to harassment or could also mean that the real progenitors were also more compact (e.g. blue compact dwarf (BCD)-like objects). This has been suggested earlier by e.g. \cite{koleva:2009} who proposed BCDs as the progenitors of dEs with steep metallicity gradients. We have added a dynamical/kinematic weight to the argument (note, however, that we find gradients in one third of our sample and the rest of our galaxies show null gradient values). While we are not in the position yet to decisively confirm or reject either hypothesis, at least the former scenario seems compatible with our latest results presented in this work.

The discovery of those oldest stellar populations underlying four galaxies in our sample and the majority of dEs studied so far provides support to the claim that they contributed to the build-up of the outskirts of massive galaxies. So far we have claimed otherwise (e.g. in \citealt{rys:2012jenam}) since massive ETGs showed older ages based on SSP estimates. The SSP ages of both types are, however, biased by the light coming from young(er) populations, if they are present. We now know this to be the case for dEs. We can thus envision a scenario in which \textit{some} of the dE \textit{progenitors} were the material used in the build-up of the halos of more massive galaxies, with the remaining gas being transformed into stars early on. Those that survived would then go through the transformation steps proposed here earlier (and elsewhere in the literature), namely undergoing later episodes of star formation induced by (among others) the tidal forces of the cluster. Naturally, detailed chemical analyses will be needed to confirm this statement, in particular we should look at chemical abundances in massive ellipticals out to several effective radii, which should ideally be complimented with these properties' redshift dependence.

\subsection{A genuinely old system?}
\label{sec-oldgal}

The analysis of one of our galaxies, VCC\,1431, reveals no evidence for stellar populations younger than $\sim$10\,Gyr, i.e. what we see in its star formation history is a uniformly old population across the entire analyzed area, in this case 1.5 effective radii. The galaxy, a nucleated dE, does not show evidence for any type of substructure such as a disk or spiral arms \citep{janz:2013}. Thus there are no tell-tale sign of interaction present. Could this object be genuinely old? 

On the one hand, following the prevailing ideas on how dE galaxies came to be, we should assume the infall scenario. The galaxy's projected distance from the Virgo's central galaxy M87 is only 1.19 degrees and thanks to the surface brightness fluctuation-based line-of sight distance \citep{mei:2007} we know that its true, intrinsic location is also a central one.  Our previous results on the dynamical properties of the galaxy (see \citealt{rys:2014}) show it to have one of the lowest angular momentum values and steepest circular velocity profile of our cluster dwarfs. We therefore conclude that the object had its SF quenched through ram-pressure stripping before it entered the cluster environment (e.g. in a group setting\footnote{a number of recent works, both observational and theoretical, suggest that ram-pressure stripping is more efficient than we used to think (see the discussion in \citealt{kormendy:2012} and the references therein); also, a good illustration of ram-pressure stripping at work in a group environment is the morphology-density relation of the Milky Way/M31 groups of satellite galaxies \citep{mateo:2008}.} and subsequent tidal influence lowered its angular momentum and heated the galaxy, making it more compact and rounder ($\epsilon=0.03$), in agreement with \cite{moore:1998}. 

On the other hand, only because a galaxy is situated close to the cluster center does not necessarily mean that something must have happened to it that left obvious traces. It may be sufficiently compact in radius that its tidal radius is well outside the baryonic one (see \citealt{penny:2009}). And it may happen to be on an orbit with little eccentricity, in which case there would be little variation of the tidal forces and thus also no need to trigger a star formation burst. Again, in $\Lambda$CDM one would expect to have a number of galaxies that have never ``fallen in'' but have instead ``been there'' (i.e. alongside the most massive progenitor) since the beginning. This seems to be another plausible evolution scenario for VCC\,1431.

In either case, the finding needs to be understood and explained in the context of what we know about the formation and evolution of dE galaxies. Is VCC\,1431 a relic that has evolved \textit{in situ} and has survived intact over (at least) the last few Gyr because of its compactness, or was the galaxy's star formation indeed quenched long ago, before it entered the cluster environment, but tidal interactions heated the body of the galaxy, altering its mass profile and/or rotational properties? One way to address these questions would be to try and find objects of similar mass and structure in cosmological simulations and trace their evolution back to before they became satellites (see Stringer et al. in prep. for such an analysis of compact giant ellipticals based on the \cite{trujillo:2014} results for NGC\,1277). This could provide more insight on the possible evolutionary paths of such objects.

\subsection{The puzzle of an isolated galaxy}

For the field galaxy ID\,0918 we see a correlation between the stellar populations radial trends and its kinematic properties. Both young and old populations have a break in their profile at around 1.2 $R_e$, which radius corresponds to the size of the kinematically-decoupled core (KDC) reported in \cite{rys:2013}. The clear positive radial age trend is broken at the KDC radius for both populations, with the ages being much lower for larger radii but again seeming to pick up a positive trend there as well. The younger population's metallicity profile is flat both inside and outside of the KDC region but the latter values are on average higher by 0.8 dex. 

This is an unusual trend which could be explained if we assume that the KDC is a result of a merger of two dwarf galaxies, with the material from the smaller accreted dwarf being the explanation for the younger ages and higher metallicities in the outskirts of ID\,0918. Because the trends are not smoothed out we could conclude that when the merger happened, both galaxies must have already lost their gas reservoirs and no further star formation took place. For this explanation to be correct, the merger must have happened within the last $\sim$1\,Gyr. Unfortunately, no deep imaging - where, e.g. tidal tails could be spotted - of the region is currently available to confirm or disprove the theory. On the other hand, the lower metallicity of the younger population in the central parts could be explained if some stellar population/material mixing has taken place.

No similar trend, i.e. no stellar population signature, is seen for the other field dwarf hosting a KDC, i.e. ID\,0650. A possible explanation for the difference could be the time at which each presumed merger occurred: long enough ago for the stellar population differences to have been diluted in ID\,0650. The availability of gas might also have induced a SF episode which smoothed the differences between the two components.

The relative isolation of the galaxies - no similarly sized or larger objects within a projected radius of 20\,arcmin - makes the suggested merger in principle a highly unlikely event. ID\,0918 must, therefore, be a result of a particular coincidence of progenitor galaxies' initial orbits which led to the suggested merger. No other theory currently exists that would be able to explain the above findings. It would, naturally, be interesting to investigate similar environments and galaxies residing in them in search of (dis)similarities with ID\,0918, and also to carry out simulations which could shed some light on its exact formation path.

\section{Summary and future work}

We have presented the stellar population analysis of a sample of 12 dwarf elliptical galaxies coming from our SAURON study of Virgo and field dE objects. We have shown that with high quality unresolved spectroscopic data  we are able to to recover star formation histories in the sense of decomposing our galaxies into two stellar populations. Additionally, by employing 3D data and using values averaged along annuli the derived properties are not biased towards a specific axis and as such are more robust than long-slit data. 

We find that the majority of our dEs either still exhibited significant star formation a few Gyr ago or they experienced a secondary burst of SF that occurred roughly at that time. We interpret this as a tentative piece of evidence in favor of a harassment scenario, where such episodes of SF are expected as the remaining gas is driven inwards as a result of tidal forces. 

We also find that the old populations of some of our dEs are roughly coeval with those of giant ETGs, in principle making it possible for the \textit{progenitors} of those dEs to have contributed to the build-up of the outskirts of massive galaxies -- the younger SSP ages of dEs simply reflecting a more extended SF in later epochs. A detailed comparison of stellar population properties of dwarf and giant early-type galaxies based on line-strength analysis is the focus of our subsequent paper (Ry\'s et al. in prep.)

We find a correlation between the stellar population and kinematic properties of the field galaxy ID\,0918. The metallicity and age profiles show a break at a radius which corresponds to the size of the KDC we earlier found in this galaxy \citep{rys:2013}. We believe it to be an argument in favor of a merger origin of the KDC.

Finally, one of our galaxies, VCC\,1431, does not show any sign of younger populations across the probed spatial extent ($\sim$1.5 effective radii). We conclude that it either was ram-pressure stripped early on in its evolution in a group environment and subsequently tidally heated (which lowered its angular momentum and increased compactness), or that it evolved \textit{in situ} in the cluster's central parts, compact enough to avoid tidal disruption.

Challenges related to observing the progenitor class of dEs are numerous.  While it is intrinsically hard to accurately assess structural parameters of high-redshift galaxies to compare them with what we see in the nearby Universe, in the case of dEs we do not even have the luxury of having candidate progenitors at higher $z$. The reason is not the uncertain evolution path which makes the progenitor search complex -- after all if a given galaxy is a relic we are simply looking for an object with characteristics matching that galaxy as closely as possible. The problem is the low masses\,/\,luminosities of dEs which virtually preclude high-z observations of objects with similar structural characteristics.

We can hope that once large samples for environments of varying density have been analyzed, more answers will emerge. It would also be of vital importance to be able to test our observational results against a larger suite of simulations. So far in works dedicated to studying the influence of tidal forces on cluster dwarfs we have dealt with remnants which were more massive (\citealt{moore:1998}) or more compact that our typical dwarfs (\citealt{smith:2010}) or have explored a limited set of initial structural parameters (\citealt{mastropietro:2005}). Ideally, one would also want to compare observations and simulations from as close a perspective as possible, by creating mock datasets (see e.g. \citealt{naab:2014}) and applying those same tools to their analysis as one does to real data. In this way our interpretation of the latter could be more robust.

\section*{Acknowledgments}

AR is a postdoctoral fellow of the Alexander von Humboldt Foundation (Germany). MK is a fellow of the Fund for Scientific Research – Flanders, Belgium (FWO). JFB acknowledges support from grant AYA2013-48226-C3-1-P from the Spanish Ministry of Economy and Competitiveness (MINECO). The paper is based on observations obtained at the William Herschel Telescope, operated by the Isaac Newton Group in the Spanish Observatorio del Roque de los Muchachos of the Instituto de Astrof\'isica de Canarias. We thank the referee, Richard McDermid, for his constructive comments and suggestions, which have helped optimize the presentation of this work. AR thanks Harald Kuntschner for comments on the draft version of this paper.

Funding for the Sloan Digital Sky Survey (SDSS) and SDSS-II has been provided by the Alfred P. Sloan Foundation, the Participating Institutions, the National Science Foundation, the U.S. Department of Energy, the National Aeronautics and Space Administration, the Japanese Monbukagakusho, the Max Planck Society, and the Higher Education Funding Council for England. The SDSS Web site is http://www.sdss.org/.

\bibliography{biblio}

\appendix

\section{Comparison of 2-SSP ULySS SFHs with line-strength-based stellar population values.}
\label{app-a}

\begin{figure}
\centering
\includegraphics[width=0.45\textwidth]{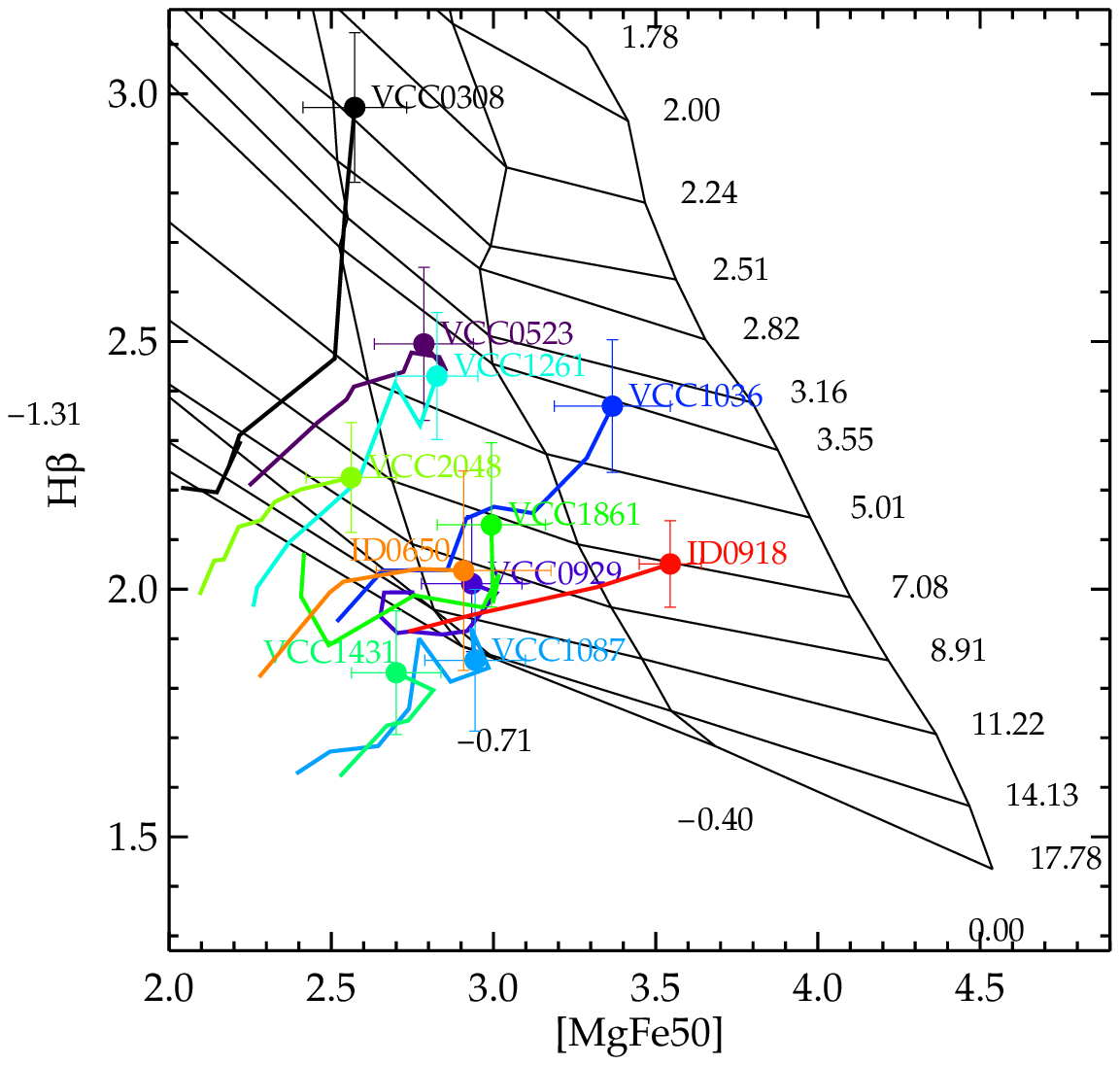}
\caption{H$\beta$ vs. [MgFe50] line strength gradients averaged along isophotes, overplotted on MILES model predictions of \protect\cite{falcon:2011a}. Galaxy centers are marked with filled dots and their error bars are also provided. The values are shown up to 1 effective radius, unless the radial extent of the data is smaller. NGC\,3073 is not shown as as it falls off the y-axis at younger ages -- the plotting range has been adjusted to emphasize the differences among the cluster galaxies.}
\label{indices} 
\end{figure}

While a detailed analysis of line strengths (LS) and LS-based stellar population parameters will be the subject of our forthcoming paper, we present here a comparison between them and the SFHs presented in this work. The LS values shown in Figure~\ref{indices} have been used to obtain ages and metallicities and their errors with the use of the $rmodel$ software of \cite{cardiel:2003}. For a detailed description we refer the reader to the above paper. In short, the program, using as input LS indices, determines ages and metallicities through the interpolation in SSP model predictions, in this case MILES models of \cite{falcon:2011a}. Both linear and bivariate fits are computed to perform the interpolation, with the latter adopted for our purposes. The errors on the parameters are estimated by running Monte-Carlo simulations, making use of the uncertainties introduced for each line-strength index.

\begin{figure*}
\centering
\hspace{1.0cm}Age \hspace{3.2cm} Metallicity \hspace{3.4cm} Age \hspace{2.9cm} Metallicity\\
\vspace{0.3cm}

\includegraphics[width=0.48\columnwidth]{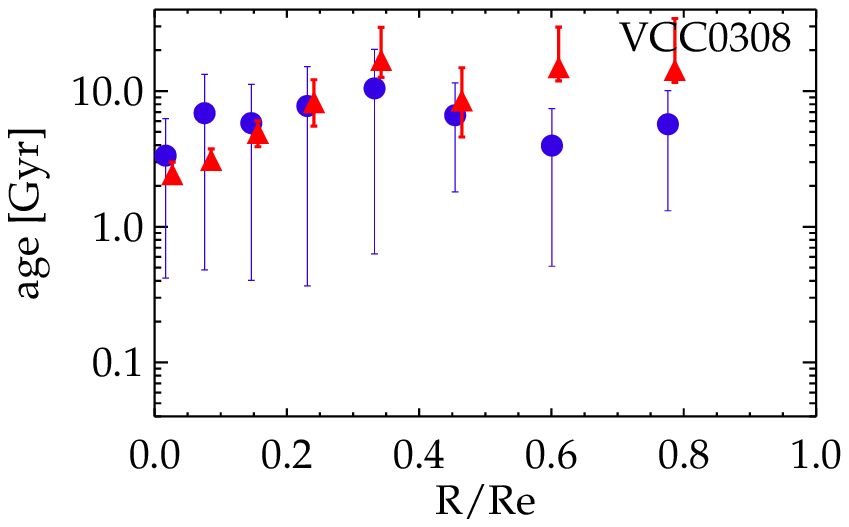}\hspace{0.2cm}
\includegraphics[width=0.48\columnwidth]{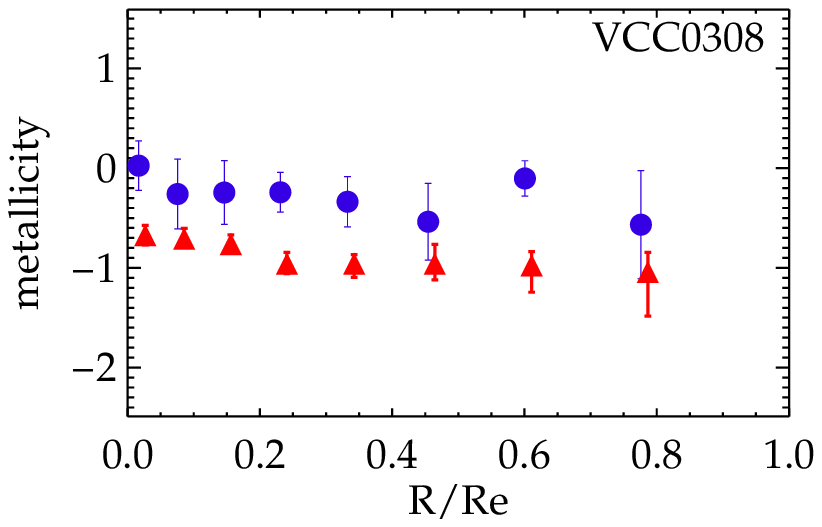}\hspace{0.8cm}
\includegraphics[width=0.48\columnwidth]{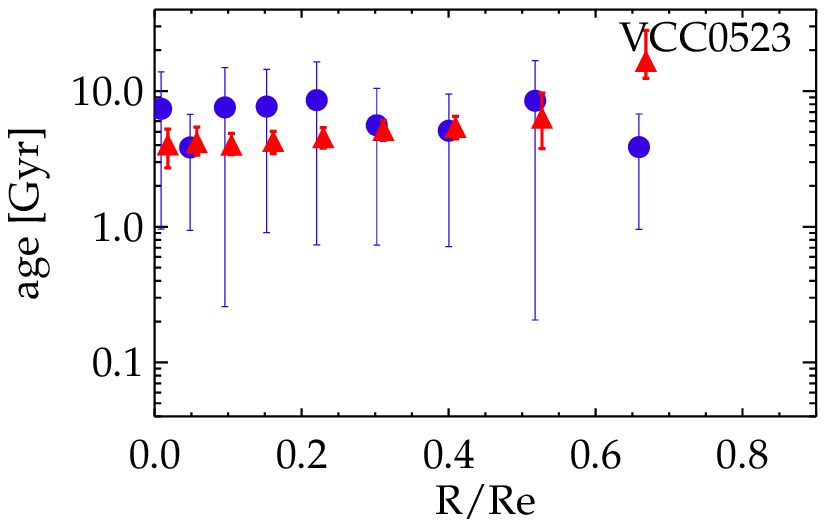}\hspace{0.2cm}
\includegraphics[width=0.48\columnwidth]{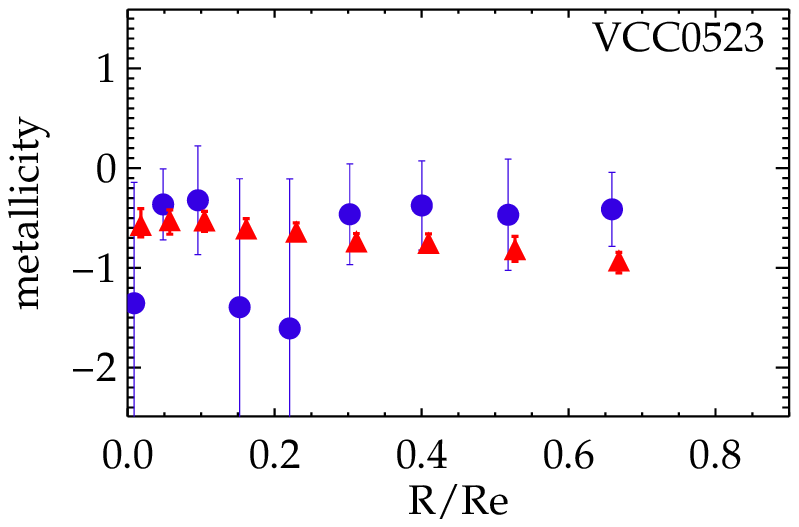}
\vspace{0.8cm}

\includegraphics[width=0.48\columnwidth]{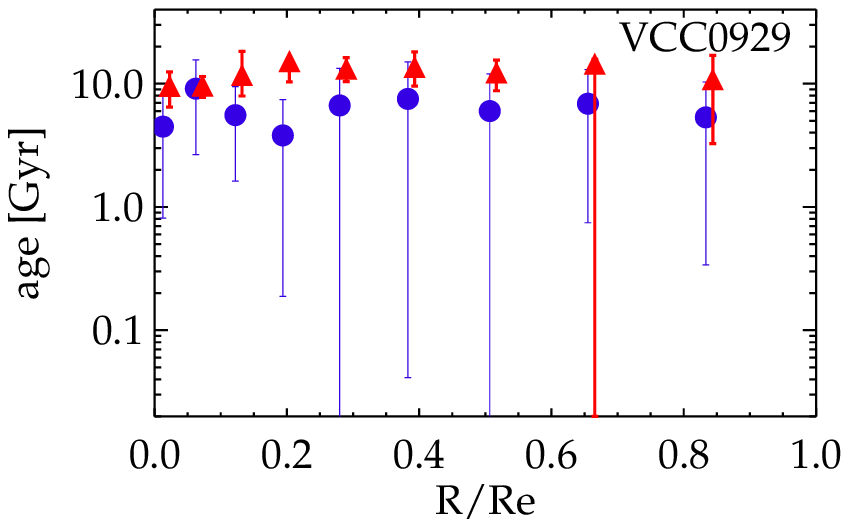}\hspace{0.2cm}
\includegraphics[width=0.48\columnwidth]{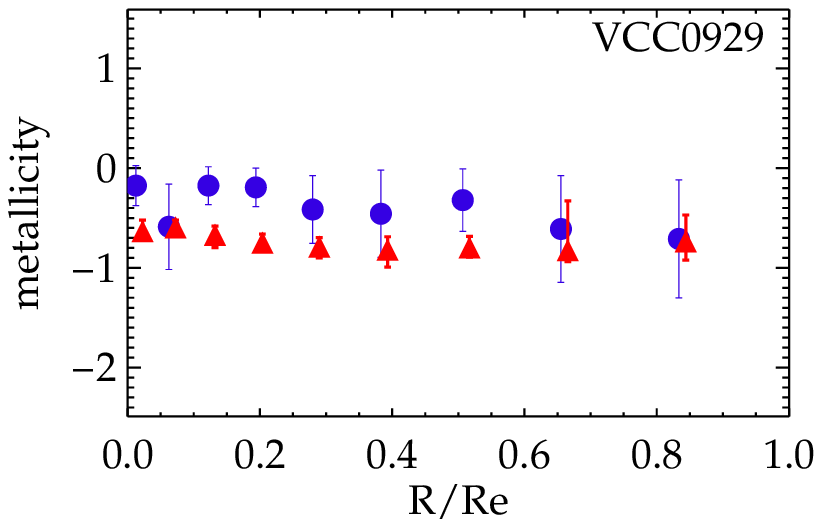}\hspace{0.8cm}
\includegraphics[width=0.48\columnwidth]{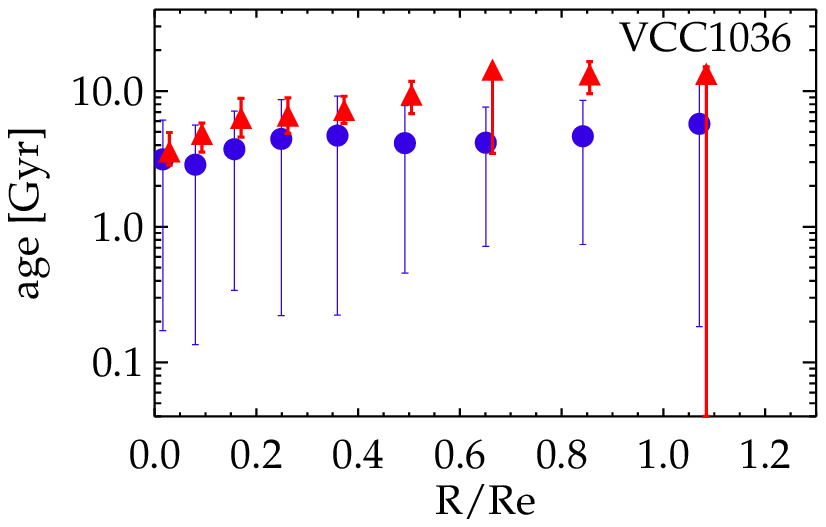}\hspace{0.2cm}
\includegraphics[width=0.48\columnwidth]{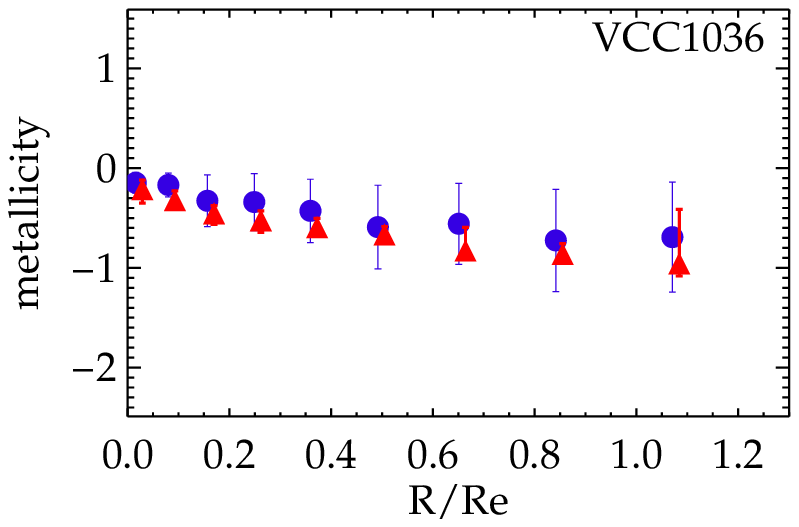}
\vspace{0.8cm}

\includegraphics[width=0.48\columnwidth]{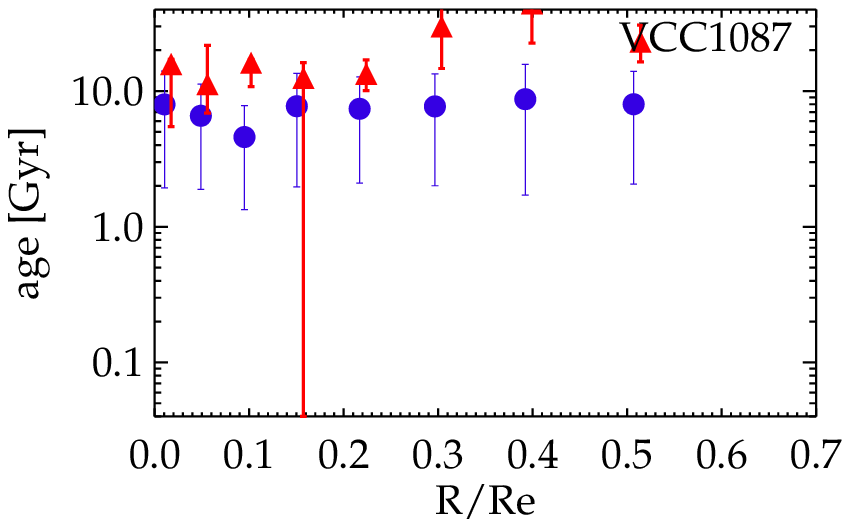}\hspace{0.2cm}
\includegraphics[width=0.48\columnwidth]{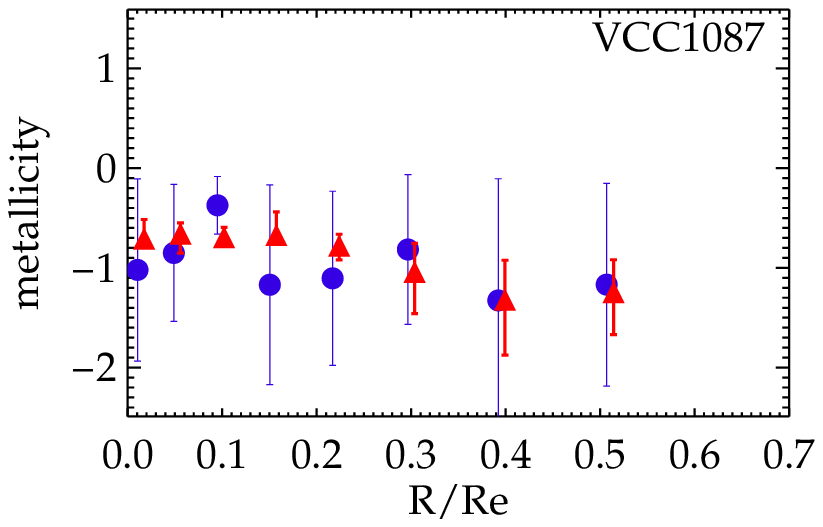}\hspace{0.8cm}
\includegraphics[width=0.48\columnwidth]{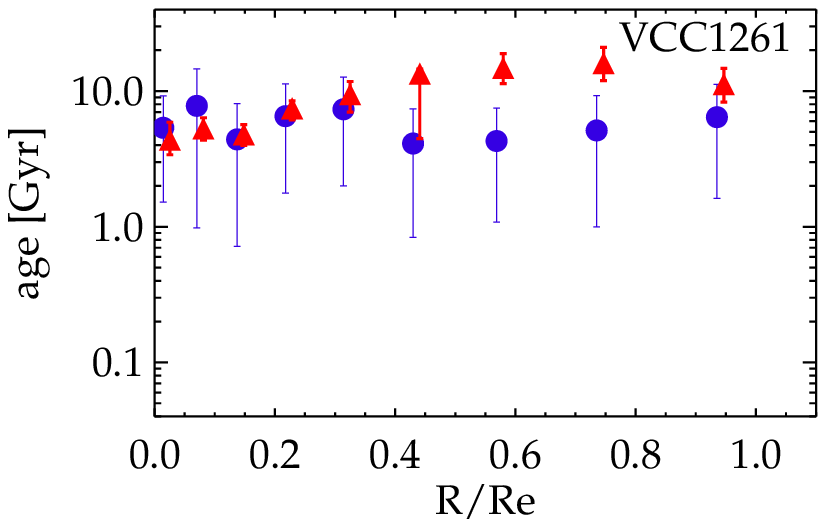}\hspace{0.2cm}
\includegraphics[width=0.48\columnwidth]{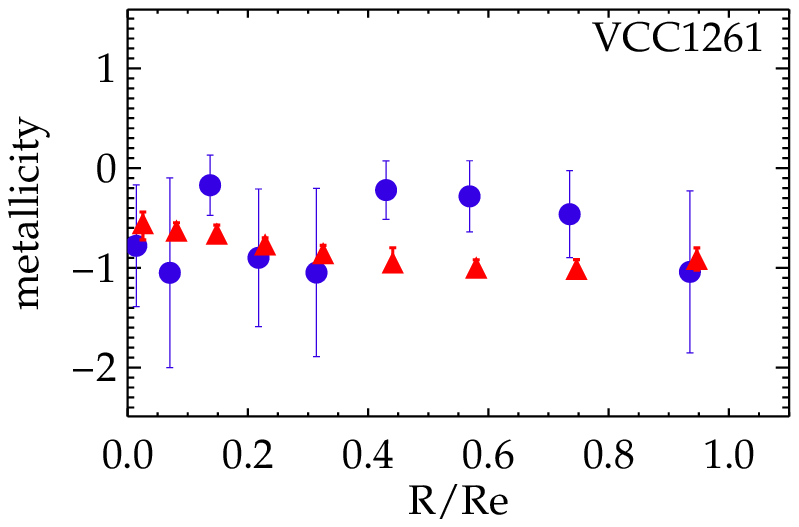}
\vspace{0.8cm}

\includegraphics[width=0.48\columnwidth]{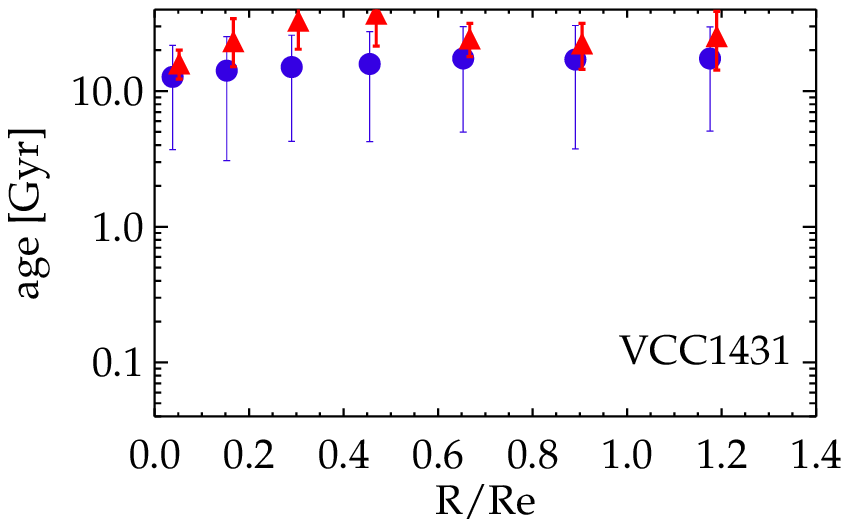}\hspace{0.2cm}
\includegraphics[width=0.48\columnwidth]{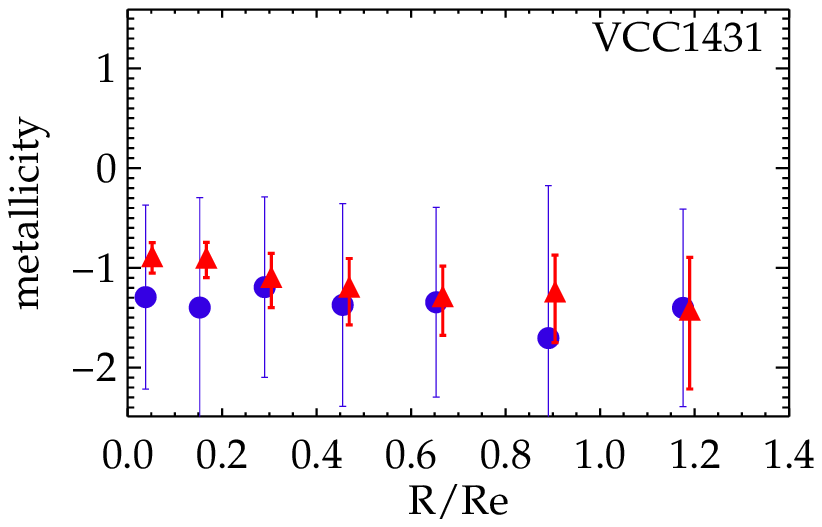}\hspace{0.8cm}
\includegraphics[width=0.48\columnwidth]{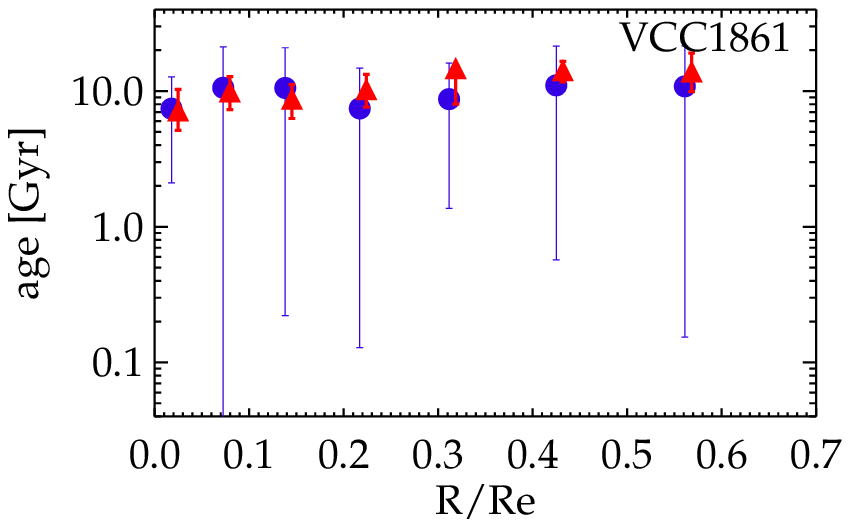}\hspace{0.2cm}
\includegraphics[width=0.48\columnwidth]{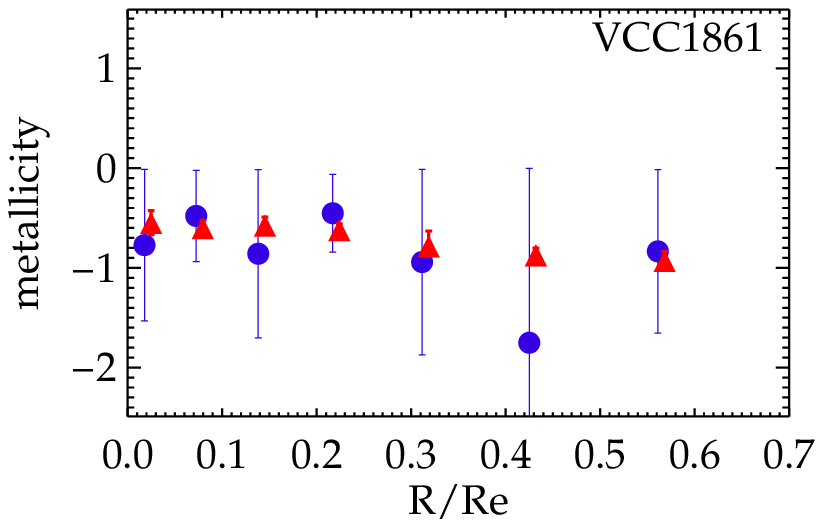}
\vspace{0.8cm}

\includegraphics[width=0.48\columnwidth]{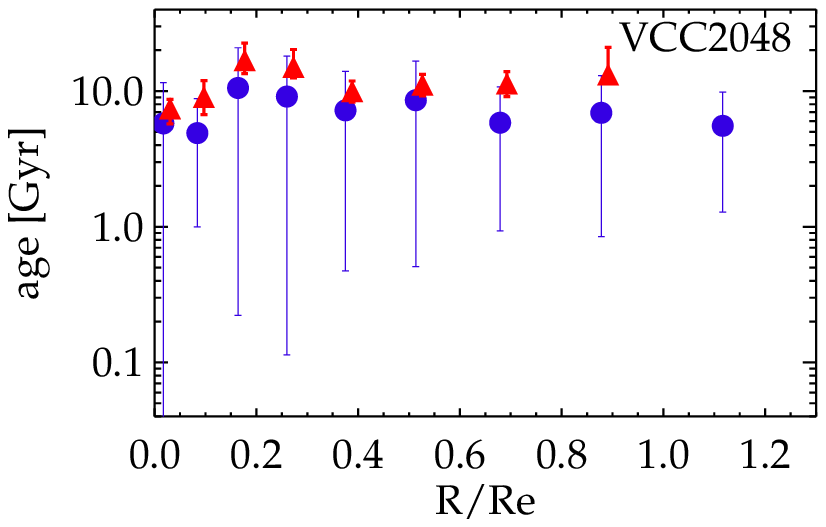}\hspace{0.2cm}
\includegraphics[width=0.48\columnwidth]{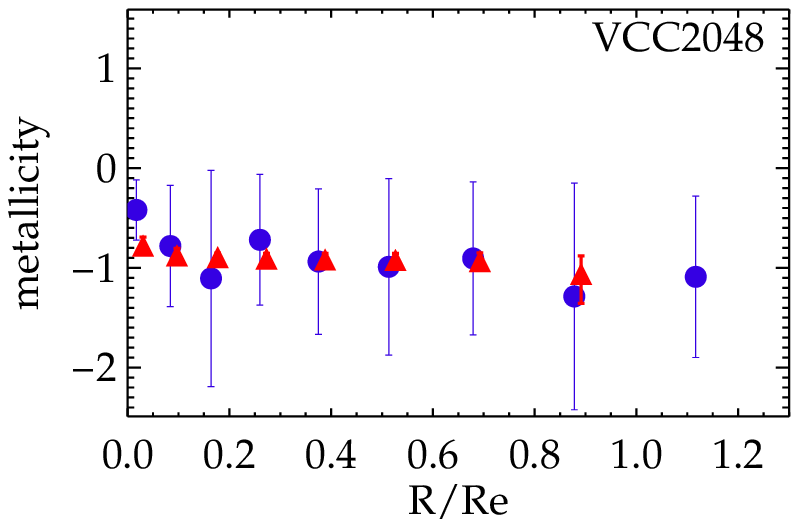}\hspace{0.8cm}
\includegraphics[width=0.48\columnwidth]{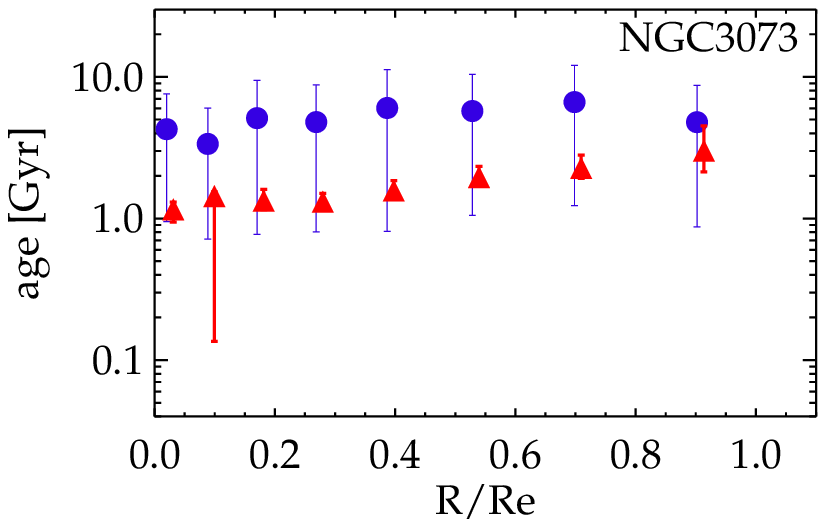}\hspace{0.2cm}
\includegraphics[width=0.48\columnwidth]{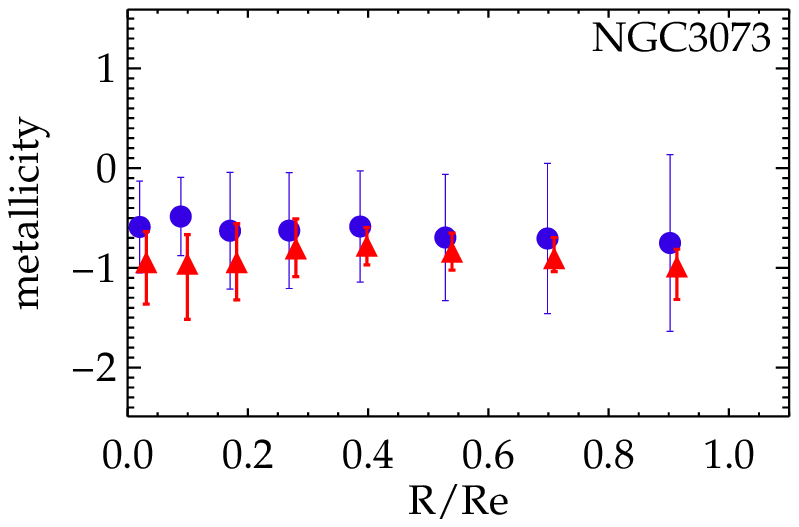}
\vspace{0.8cm}

\includegraphics[width=0.48\columnwidth]{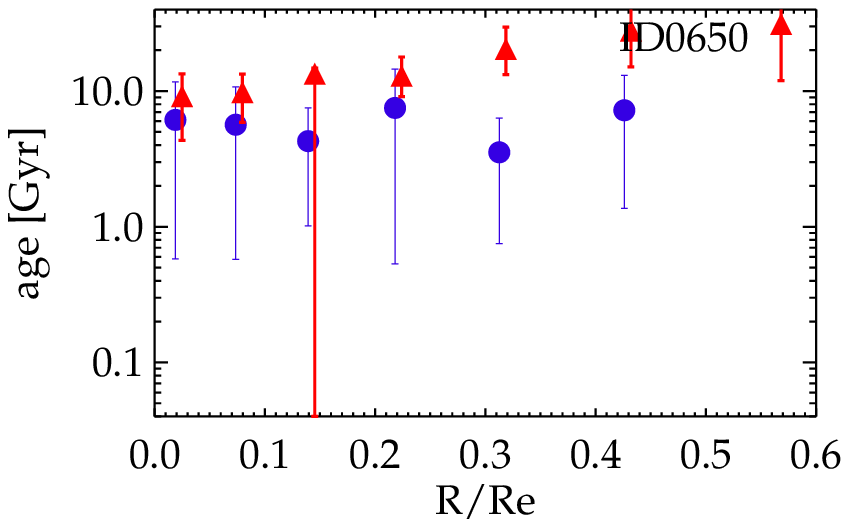}\hspace{0.2cm}
\includegraphics[width=0.48\columnwidth]{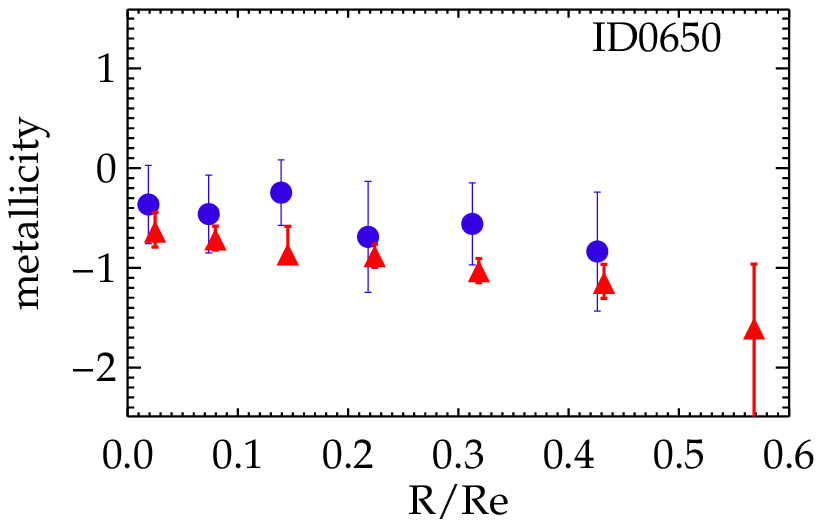}\hspace{0.8cm}
\includegraphics[width=0.48\columnwidth]{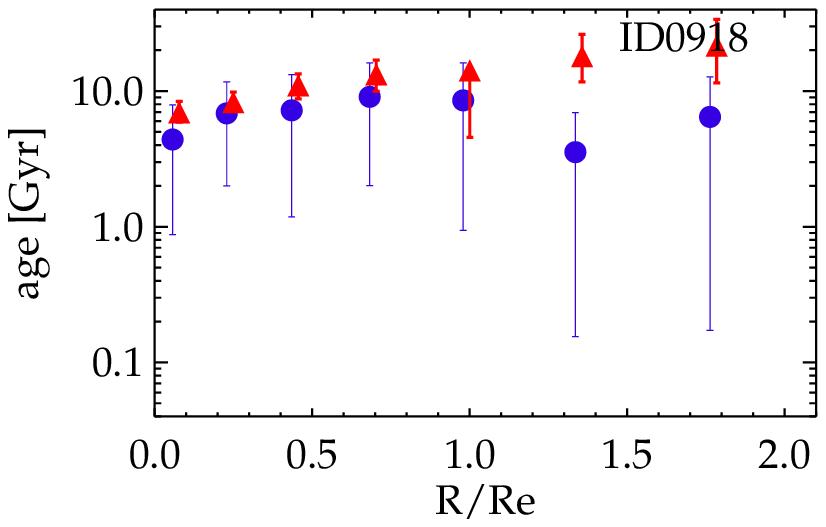}\hspace{0.2cm}
\includegraphics[width=0.48\columnwidth]{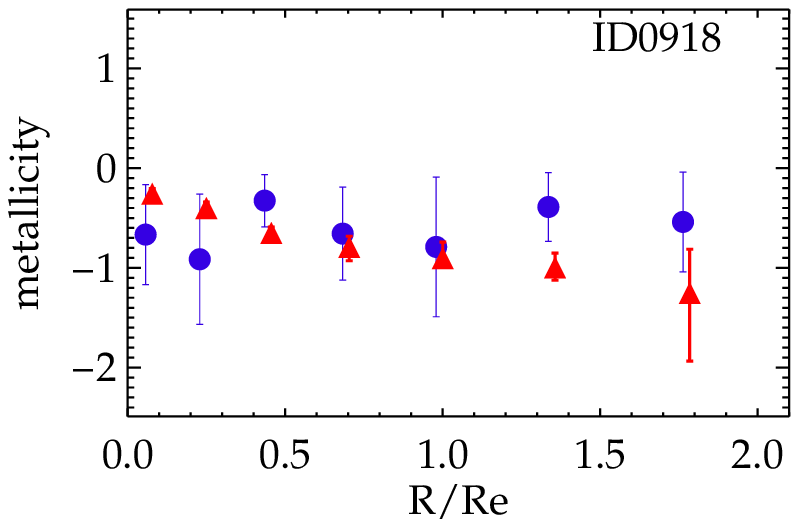}
\vspace{0.8cm}

\caption{Comparison of luminosity weighted profiles from the SFHs presented in this work (blue filled circles) with the profiles of SSP age and $Z$ values from our line-strength analysis (red filled triangles) to be presented in Ry\'s et al. (in prep.).}
\label{sfh-profiles_LS}  
\end{figure*}

This comparison is intended as a consistency check between the two independent methods, also given the complex nature of multiple-SSP approach applied to unresolved data. Figure~\ref{sfh-profiles_LS} shows that for most of the galaxies we find a good agreement between the SFH luminosity-weighted and LS-based ages and metallicities, particularly in the innermost (typically younger) regions. Below we comment on those objects for which some systematic deviations are found, noting, however, that most of them still agree to within the errors. Our overall conclusion is that nearly consistent results are obtained for most galaxies.

For VCC0\,308 the LS results suggest lower metallicity, which does  not translate into an older population in the inner region (i.e against the age/metallicity degeneracy expectations). In the case of VCC\,0929 there is a systematic difference in age, which is compensated by a difference in metallicity. For VCC\,1087 the agreement is good given the quite old ages, however, some disagreement is seen in the very center, and the age/metallicity degeneracy might be at work. VCC\,1261 innermost regions show lower LS metallicities, while ages are in good agreement. In the case of ID\,0918 the SFH metallicities in the center are lower than the LS-based values, with an opposite trend in the outer parts.

\section{ULySS-extracted population parameters}
\label{app-b}

Monte-Carlo (MC) simulations are performed to estimate errors, and the coupling (i.e. degeneracies) between the parameters. Table~\ref{ulyssmc} shows the ULySS-extracted SFHs including MC simulation results for all radial bins and galaxies. The simulations consist of series of analyses of a spectrum with added random noise corresponding to the noise estimated from the data (200 realizations in each case), computed to produce $\chi^2$ close to 1. The added noise has a Gaussian distribution and takes into account the correlation between the pixels introduced along the processing. This effect is modeled by keeping track of the number of independent pixels during the steps of the processing, and then generating a random vector of independent points which is eventually resampled to the actual length of the spectrum (see section 2 and 3.3 of \citealt{koleva:2009} for more details).


\centering
\onecolumn

\begin{deluxetable}{ccccc}
\tabletypesize{\small}
\tablewidth{0pt}
\tablecolumns{5}
\tablecaption{Results of the ULySS Monte-Carlo simulations.\label{ulyssmc}}
\tablehead{
 \colhead{r} &
 \colhead{age (young)} &
 \colhead{Z (young)} &
 \colhead{age (old)} &
 \colhead{Z (old)} \\
 \colhead{(arcsec)} &
 \colhead{(Myr)} &
 \colhead{(Z$_\odot$)} &
 \colhead{Myr} &
 \colhead{(Z$_\odot$)}
}
\startdata
\multicolumn{5}{c}{VCC0308}\\
\hline
      1.00000&795$\pm$610&+0.33$\pm$0.43&15215$\pm$2530&-0.86$\pm$0.49\\
      2.20000&1433$\pm$2826&+0.23$\pm$0.54&15173$\pm$3413&-0.81$\pm$0.65\\
      3.64000&1233$\pm$1907&+0.20$\pm$0.38&15849$\pm$1981&-0.92$\pm$0.52\\
      5.37000&1881$\pm$2593&-0.25$\pm$0.60&14213$\pm$3962&-0.37$\pm$0.81\\
      7.44000&1676$\pm$2124&-0.28$\pm$0.64&16365$\pm$2638&-0.38$\pm$0.82\\
      9.93000&3543$\pm$5323&-0.43$\pm$0.73&14730$\pm$5597&-0.83$\pm$0.77\\
      12.9100&3484$\pm$4116&-1.04$\pm$0.54&13635$\pm$4080&+0.25$\pm$0.85\\
      16.4900&3690$\pm$4107&-1.23$\pm$1.05&14052$\pm$3940&-0.09$\pm$0.87\\
\hline
\multicolumn{5}{c}{VCC0523}\\
\hline
      1.00000&690$\pm$74&+0.50$\pm$0.07&11999$\pm$0&-0.84$\pm$0.03\\
      2.20000&1354$\pm$708&+0.07$\pm$0.22&11999$\pm$0&-1.26$\pm$0.69\\
      3.64000&1265$\pm$672&+0.11$\pm$0.20&11999$\pm$0&-1.23$\pm$0.68\\
      5.37000&694$\pm$140&+0.28$\pm$0.07&11999$\pm$0&-0.80$\pm$0.13\\
      7.44000&1561$\pm$740&-0.01$\pm$0.22&11999$\pm$0&-1.62$\pm$0.74\\
      9.93000&1615$\pm$738&-0.06$\pm$0.23&11999$\pm$0&-1.67$\pm$0.71\\
      12.9100&2268$\pm$87&-0.26$\pm$0.02&11999$\pm$0&-2.29$\pm$0.10\\
      16.4900&2276$\pm$26&-0.33$\pm$0.02&11999$\pm$0&-2.30$\pm$0.01\\
      20.7900&2276$\pm$65&-0.33$\pm$0.03&11999$\pm$0&-2.27$\pm$0.05\\
\hline
\multicolumn{5}{c}{VCC0929}\\
\hline
      1.00000&1985$\pm$3246&+0.05$\pm$0.58&6959$\pm$2570&-0.39$\pm$0.57\\
      2.20000&7919$\pm$4134&-0.68$\pm$0.68&10731$\pm$4661&-0.49$\pm$0.63\\
      3.64000&4237$\pm$2935&+0.02$\pm$0.65&8769$\pm$1786&-0.63$\pm$0.50\\
      5.37000&365$\pm$1345&-0.00$\pm$0.63&7850$\pm$1033&-0.42$\pm$0.16\\
      7.44000&28$\pm$19&-0.27$\pm$0.54&9921$\pm$871&-0.49$\pm$0.03\\
      9.93000&197$\pm$727&-0.10$\pm$0.35&9588$\pm$591&-0.56$\pm$0.08\\
      12.9100&18$\pm$8&-0.02$\pm$0.15&10151$\pm$617&-0.53$\pm$0.02\\
      16.4900&2412$\pm$1756&-0.25$\pm$0.74&9212$\pm$2659&-0.80$\pm$0.40\\
      20.7900&1406$\pm$1499&-0.53$\pm$0.57&6729$\pm$2738&-0.78$\pm$0.50\\
\hline
\multicolumn{5}{c}{VCC1036}\\
\hline
      1.00000&505$\pm$840&-0.24$\pm$0.32&4705$\pm$3322&-0.10$\pm$0.36\\
      2.20000&463$\pm$699&-0.31$\pm$0.25&3956$\pm$532&-0.11$\pm$0.15\\
      3.64000&2116$\pm$1866&-0.48$\pm$0.24&4110$\pm$1956&-0.30$\pm$0.22\\
      5.37000&1341$\pm$2294&-0.36$\pm$0.21&5138$\pm$1960&-0.34$\pm$0.16\\
      7.44000&716$\pm$536&-0.46$\pm$0.16&6677$\pm$1612&-0.42$\pm$0.04\\
      9.93000&1219$\pm$1002&-0.80$\pm$0.21&6184$\pm$964&-0.46$\pm$0.12\\
      12.9100&1665$\pm$1378&-0.71$\pm$0.34&6696$\pm$1952&-0.42$\pm$0.26\\
      16.4900&1927$\pm$1401&-0.88$\pm$0.19&6944$\pm$1490&-0.63$\pm$0.15\\
      20.7900&1037$\pm$1092&-0.92$\pm$0.39&6852$\pm$1380&-0.65$\pm$0.10\\
\hline
\multicolumn{5}{c}{VCC1087}\\
\hline
      1.00000&2767$\pm$2170&-0.13$\pm$0.31&11999$\pm$0&-1.35$\pm$0.94\\
      2.20000&4688$\pm$1033&-0.12$\pm$0.20&11999$\pm$0&-1.78$\pm$0.43\\
      3.64000&4731$\pm$753&-0.22$\pm$0.20&11999$\pm$0&-1.97$\pm$0.51\\
      5.37000&4654$\pm$1261&-0.26$\pm$0.08&11999$\pm$0&-2.17$\pm$0.46\\
      7.44000&4951$\pm$479&-0.30$\pm$0.02&11999$\pm$0&-2.28$\pm$0.17\\
      9.93000&4201$\pm$1321&-0.28$\pm$0.12&11999$\pm$0&-2.13$\pm$0.44\\
      12.9100&4604$\pm$658&-0.33$\pm$0.03&11999$\pm$0&-2.29$\pm$0.12\\
      16.4900&4896$\pm$340&-0.44$\pm$0.04&11999$\pm$0&-2.30$\pm$0.01\\
\hline
\multicolumn{5}{c}{VCC1261}\\
\hline
      1.00000&1168$\pm$605&+0.16$\pm$0.32&11999$\pm$0&-0.94$\pm$0.44\\
      2.20000&1354$\pm$535&+0.09$\pm$0.15&11999$\pm$0&-1.10$\pm$0.46\\
      3.64000&1876$\pm$914&+0.09$\pm$0.22&11999$\pm$0&-1.46$\pm$0.49\\
      5.37000&1770$\pm$958&-0.04$\pm$0.26&11999$\pm$0&-1.32$\pm$0.55\\
      7.44000&2473$\pm$1080&-0.24$\pm$0.27&11999$\pm$0&-1.69$\pm$0.55\\
      9.93000&3340$\pm$578&-0.44$\pm$0.09&11999$\pm$0&-2.12$\pm$0.28\\
      12.9100&3775$\pm$1553&-0.40$\pm$0.22&11999$\pm$0&-1.88$\pm$0.53\\
      16.4900&4566$\pm$921&-0.39$\pm$0.16&11999$\pm$0&-2.13$\pm$0.34\\
      20.7900&3075$\pm$688&-0.28$\pm$0.12&11999$\pm$0&-2.21$\pm$0.29\\
\hline
\multicolumn{5}{c}{VCC1431}\\
\hline
      1.00000&10797$\pm$4528&-1.32$\pm$0.95&15412$\pm$2680&-1.29$\pm$0.90\\
      2.20000&11595$\pm$3276&-1.11$\pm$0.90&15812$\pm$2857&-1.58$\pm$0.90\\
      3.64000&15073$\pm$1616&-0.63$\pm$0.61&15729$\pm$3549&-2.07$\pm$0.62\\
      5.37000&14835$\pm$3608&-1.21$\pm$0.92&16758$\pm$1672&-1.50$\pm$0.91\\
      7.44000&17477$\pm$1853&-1.50$\pm$0.91&17648$\pm$619&-1.25$\pm$0.90\\
      9.93000&15722$\pm$1716&-0.64$\pm$0.46&17895$\pm$439&-2.17$\pm$0.46\\
      12.9100&17202$\pm$1861&-1.51$\pm$0.87&17876$\pm$378&-1.33$\pm$0.88\\
\hline
\multicolumn{5}{c}{VCC1861}\\
\hline
      1.00000&1721$\pm$1739&-0.12$\pm$0.25&11999$\pm$0&-0.98$\pm$0.81\\
      2.20000&3910$\pm$1830&-0.03$\pm$0.16&11999$\pm$0&-1.55$\pm$0.64\\
      3.64000&4893$\pm$579&-0.02$\pm$0.09&11999$\pm$0&-2.01$\pm$0.22\\
      5.37000&4999$\pm$0&-0.02$\pm$0.03&11999$\pm$0&-2.11$\pm$0.06\\
      7.44000&1549$\pm$2122&-0.16$\pm$0.21&11999$\pm$0&-0.92$\pm$0.84\\
      9.93000&198$\pm$106&-0.20$\pm$0.14&11999$\pm$0&-0.52$\pm$0.03\\
      12.9100&342$\pm$787&-0.18$\pm$0.15&11999$\pm$0&-0.63$\pm$0.27\\
\hline
\multicolumn{5}{c}{VCC2048}\\
\hline
      1.00000&1727$\pm$566&-0.17$\pm$0.24&11999$\pm$0&-1.26$\pm$0.29\\
      2.20000&1642$\pm$680&-0.35$\pm$0.22&11999$\pm$0&-1.31$\pm$0.54\\
      3.64000&1773$\pm$231&-0.26$\pm$0.09&11999$\pm$0&-1.88$\pm$0.34\\
      5.37000&1869$\pm$107&-0.33$\pm$0.04&11999$\pm$0&-2.25$\pm$0.12\\
      7.44000&1741$\pm$401&-0.31$\pm$0.15&11999$\pm$0&-1.84$\pm$0.38\\
      9.93000&2439$\pm$464&-0.57$\pm$0.05&11999$\pm$0&-2.18$\pm$0.39\\
      12.9100&497$\pm$1020&-0.29$\pm$0.31&11999$\pm$0&-0.87$\pm$0.60\\
      16.4900&1141$\pm$1380&-0.59$\pm$0.29&11999$\pm$0&-1.22$\pm$0.74\\
      20.7900&36$\pm$13&-0.34$\pm$0.36&11999$\pm$0&-0.52$\pm$0.10\\
\hline
\multicolumn{5}{c}{ID0650}\\
\hline
      1.00000&1179$\pm$379&+0.05$\pm$0.23&11999$\pm$0&-0.85$\pm$0.27\\
      2.20000&1094$\pm$861&-0.14$\pm$0.57&11999$\pm$0&-0.91$\pm$0.42\\
      3.64000&1726$\pm$683&+0.10$\pm$0.22&11999$\pm$0&-1.28$\pm$0.42\\
      5.37000&1388$\pm$1151&-0.39$\pm$0.48&11999$\pm$0&-0.92$\pm$0.49\\
      7.44000&1183$\pm$1260&-0.47$\pm$0.54&11999$\pm$0&-0.92$\pm$0.35\\
      9.93000&3103$\pm$998&-0.93$\pm$0.15&11999$\pm$0&-0.77$\pm$0.40\\
\hline
\multicolumn{5}{c}{ID0918}\\
\hline
      1.00000&2283$\pm$2407&-1.01$\pm$0.81&6360$\pm$2312&-0.40$\pm$0.81\\
      2.20000&5744$\pm$3858&-0.92$\pm$0.92&8727$\pm$4521&-0.95$\pm$1.06\\
      3.64000&4290$\pm$2178&-0.81$\pm$0.39&8655$\pm$2772&-0.11$\pm$0.53\\
      5.37000&6454$\pm$1810&-0.91$\pm$0.71&10849$\pm$3393&-0.51$\pm$0.85\\
      7.44000&2798$\pm$3916&-0.27$\pm$0.57&11961$\pm$5520&-1.10$\pm$0.65\\
      9.93000&330$\pm$898&-0.10$\pm$0.13&6646$\pm$2694&-0.67$\pm$0.36\\
      12.9100&499$\pm$1566&-0.12$\pm$0.28&9964$\pm$2759&-0.79$\pm$0.44\\
\enddata
\end{deluxetable}

\addtocounter{table}{-1}

\begin{deluxetable}{ccccccc}
\tabletypesize{\small}
\tablewidth{0pt}
\tablecolumns{7}
\tablecaption{(cont'd) \label{ulyssmc2}}
\tablehead{
 \colhead{r} &
 \colhead{age (young)} &
 \colhead{Z (young)} &
  \colhead{age (interm.)} &
 \colhead{Z (interm.)} &
 \colhead{age (old)} &
 \colhead{Z (old)} \\
 \colhead{(arcsec)} &
 \colhead{(Myr)} &
 \colhead{(Z$_\odot$)} &
  \colhead{(Myr)} &
 \colhead{(Z$_\odot$)} &
 \colhead{Myr} &
 \colhead{(Z$_\odot$)}
}
\startdata

\multicolumn{7}{c}{NGC3073}\\
\hline
      1.00000&81$\pm$114&-0.15$\pm$0.34&2398$\pm$5246&-0.21$\pm$0.88&13491$\pm$6075&-1.93$\pm$0.84\\
      2.20000&375$\pm$84&-0.02$\pm$0.25&2865$\pm$5447&-0.46$\pm$0.90&11607$\pm$6324&-1.72$\pm$0.90\\
      3.64000&379$\pm$250&+0.22$\pm$0.21&2225$\pm$3114&-0.60$\pm$0.53&17171$\pm$3080&-2.20$\pm$0.30\\
      5.37000&443$\pm$1257&+0.39$\pm$0.24&1691$\pm$1174&-0.62$\pm$0.20&17826$\pm$1636&-2.27$\pm$0.26\\
      7.44000&1098$\pm$2877&-0.03$\pm$0.37&1340$\pm$554&-0.05$\pm$0.18&17095$\pm$2812&-1.84$\pm$0.34\\
      9.93000&742$\pm$1985&+0.54$\pm$0.28&1932$\pm$1664&-0.44$\pm$0.29&17308$\pm$3029&-2.19$\pm$0.45\\
      12.9100&1515$\pm$2985&+0.45$\pm$0.33&2776$\pm$2783&-0.10$\pm$0.47&12316$\pm$5707&-1.74$\pm$0.57\\
      16.4900&2194$\pm$2750&+0.42$\pm$0.37&3026$\pm$3149&+0.33$\pm$0.48&5991$\pm$2572&-1.37$\pm$0.29\\
      20.7900&2188$\pm$4513&+0.14$\pm$0.32&9673$\pm$5771&+0.04$\pm$0.65&7691$\pm$2440&-1.46$\pm$0.70\\

\enddata
\end{deluxetable}


\end{document}